\documentclass[final,5p,times,twocolumn]{elsarticle}
\usepackage{graphicx}
\usepackage{dcolumn}
\usepackage{bm}
\usepackage{longtable}
\usepackage{lscape}
\usepackage{multirow}
\usepackage{booktabs}
\usepackage{float}
\usepackage{chemmacros}
\usepackage{rotating}
\usepackage{caption}
\usepackage{placeins}
\usepackage{multicol}
\usepackage{xcolor}
\usepackage[colorlinks=true,linkcolor=blue,citecolor=blue,urlcolor=blue]{hyperref}
\usepackage{miller}
\usepackage{amsmath} 

\journal{Journal of Magnesium and Alloys}

\begin{document}

\begin{frontmatter}

\title{Defect-controlled twin activation in crystallographically equivalent magnesium micropillars}

\author[1]{Hexin Wang\corref{cor1}}

\ead{wang@imm.rwth-aachen.de}
\affiliation[1]{organization={Institute of Physical Metallurgy and Materials Physics, RWTH Aachen University},
    city={52056 Aachen},
    country={Germany}}
\author[2]{Fatim Zahra Mouhib}
\author[1]{Chunhua Tian}
\author[1]{Sang-Hyeok Lee}
\author[3]{Henry Ovri}
\author[2]{Julien Guénolé}

\affiliation[2]{organization={CNRS, Université de Lorraine, Arts et Métiers Institute of Technology, LEM3},
    city={57070 Metz},
    country={France}}
    
\affiliation[3]{organization={Institute of Hydrogen Technology, Helmholtz Zentrum Hereon},
    city={21502 Geesthacht},
    country={Germany}}
\author[1]{Sandra Korte-Kerzel}
\author[1]{Talal Al-Samman\corref{cor1}} 
\cortext[cor1]{Corresponding author}
\ead{tasamman@imm.rwth-aachen.de}
\author[1]{Zhuocheng Xie\corref{cor1}}
\ead{xie@imm.rwth-aachen.de}

\begin{abstract}
Tensile twinning plays a central role in accommodating \hkl<c>-axis plasticity in Mg. In bulk Mg, it typically shows a relatively deterministic response with a low critical stress, whereas in confined volumes it exhibits broad yield-stress distributions that complicate the prediction of small-scale mechanical behavior. Here, site-specific compression tests are performed on 4 µm-diameter pillars fabricated in a parent Mg crystal and an adjacent \hkl{10-12} twin. The two regions share the same \hkl[11-20] compression axis but experienced different prior deformation histories, allowing the influence of the residual microstructural state to be examined at fixed crystallographic orientation. Among 27 pillars, most parent-region pillars yield near 300 MPa, whereas pillars from the twin region span approximately 30 to 300 MPa. Interrupted tests combined with cross-sectional EBSD link individual load drops to discrete twin formation and further show that a pillar containing a pre-existing twin yields at approximately 80 MPa through the migration of the existing twin boundary. Molecular dynamics simulations of 30 nm-diameter pillars resolve possible atomistic pathways at the nanoscale. The simulations illustrate how contact geometry and pre-existing twin embryos alter event selection, and how an activated twin advances rapidly, while coherent twin boundary migration proceeds through disconnection motion accompanied by crystallographically required atomic shuffles. The results attribute the experimental scatter to the
local availability of embryos and mobile interfaces, such that the first plastic event is governed by the twinning pathway accessible from the local microstructural state rather than by a single characteristic critical stress. Deformation history can therefore strongly modify the distribution of first plastic events even when the loading orientation is fixed. 

\end{abstract}

\end{frontmatter}
\section{Introduction}

Magnesium (Mg) and its alloys have garnered significant interest as energy-efficient, lightweight materials for crucial structural applications spanning the automotive, aerospace, electronics, and biomedical fields. This interest is primarily due to their superior properties, such as excellent dimensional stability, high specific strength, relative abundance, and good recyclability \cite{pollock2010weight,nie2020microstructure,xie2025atomic}. However, their broader adoption remains limited by low room-temperature formability due to the hexagonal close-packed (HCP) crystal structure, which restricts the number of available independent slip systems. In particular, the pronounced disparity in the critical resolved shear stress (CRSS) between basal and non-basal slip modes, such as prismatic \hkl<c> slip and pyramidal \hkl<c+a> slip, suppresses dislocation-mediated deformation along the \hkl<c>-axis at ambient conditions \cite{conrad1957effect,reed1957deformation,flynn1961thermally,ando199211,ovri2023mechanistic}. As a result, plastic deformation along the \hkl<c>-axis is predominantly accommodated by \hkl{10-12} tensile twinning, which has a comparatively low CRSS and plays a central role in governing the plastic response of Mg and its alloys \cite{kelley1968plane,christian1995deformation,tome2011multi}.

Historically, the fundamental principles of deformation twinning have been established through studies on bulk polycrystalline Mg, where the macroscopic mechanical response represents the statistical average of numerous twinning events \cite{christian1995deformation, yoo1981slip,godet2006use,beyerlein2010statistical}. An abundance of potential nucleation sites exists in such systems and the collective behavior of grains gives rise to relatively deterministic twinning responses that can often be rationalized using macroscopic criteria such as Schmid's law \cite{godet2006use}. For example, the CRSS for tensile twinning in bulk Mg is approximately 7–10 MPa, with only weak temperature dependence over the range of 150-450$^{\circ}$C \cite{chapuis2011temperature}. In contrast, when the characteristic dimensions are reduced to the micrometer and sub-micrometer regimes, deformation behavior departs markedly from bulk expectations, depicting the well-documented "smaller is stronger" size effect with an inverse dependence of strength on micropillar volume \cite{uchic2004sample, yu2010strong}.

In Mg micropillars, this is manifested by an increase in CRSS for twinning. The size effect is most pronounced in small pillars and becomes negligible for pillars with diameters above 10 µm. \cite{prasad2014micropillar,liu2017experimentally,wang2020effect,della2021101}.
Moreover, a size-dependent transition in deformation modes has been reported \cite{liu2017experimentally,sim2018anomalous}. Under \hkl<a>-axis compression, larger pillars exhibit nucleation and interaction of multiple twin variants, whereas smaller pillars (below 18 µm) are frequently dominated by the propagation of a single twin \cite{sim2018anomalous}. At even smaller scales (approximately 5 µm), deformation is characterized by pronounced strain bursts and significant scatter in critical stresses \cite{sim2019effect}. These observations point to a breakdown of deterministic descriptions of plasticity and suggest that twinning in confined volumes, analogously to purely dislocation-driven plasticity, becomes governed by stochastic, discrete events, strongly influenced by random local stress fluctuations and defect configurations \cite{bei2008effects,kraft2010plasticity,greer2011plasticity}.

\begin{figure*}[h!]
\centering
\includegraphics[width=\textwidth]{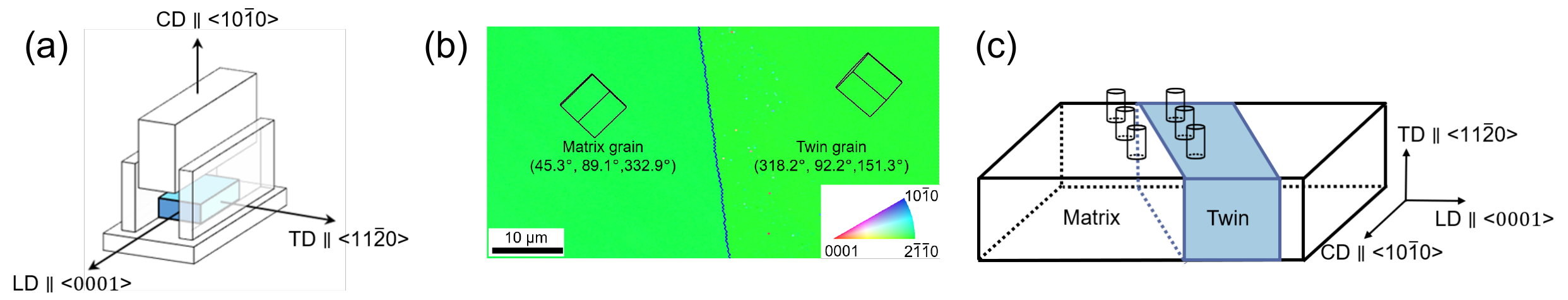}
\caption{Experimental pre-loading setup: (a) Schematic illustration of the channel-die configuration with the compression direction (CD) aligned with the \hkl[10-10] direction of the monocrystalline sample. (b) Inverse pole figure (IPF)-TD map of the region of interest (ROI) from which the micropillars were fabricated. The corresponding orientations of the matrix and the twin along with the respective Bunge Euler angles are also presented. (c) Schematic illustration of the fabricated micropillars and their location with respect to the matrix and twin grains in the pre-deformed sample.}
\label{fig_Experimental_Setup}
\end{figure*}

Extensive investigations of small-scale Mg systems have established that \hkl{10-12} tensile twins nucleate at local stress concentrators, advance rapidly through the parent crystal, and thicken through migration of their bounding interfaces  \cite{jeong2018situ,della2023micromechanical,della2023temperature,mathis2021dynamics,jeong2025nanoscale}. However, it remains unresolved how the availability of a pre-existing embryo or mobile twin boundary (TB) governs the large scatter in twinning onset stresses observed at small scales when the crystallographic loading orientation is held fixed. While prior studies have suggested that variations in pre-existing dislocation sources may contribute to this scatter \cite{sim2019effect}, a direct and systematic assessment of defect-controlled twinning under geometrically and crystallographically comparable conditions is still lacking.

In this study, we address this question by comparing site-specific micropillars fabricated in a parent crystal and an adjacent \hkl{10-12} twin. Both regions are compressed along the \hkl<a>-axis, but they have experienced different deformation histories and retain different residual microstructural states. Interrupted compression and cross-sectional EBSD are used to associate discrete load drops with the twins present after individual events. Complementary molecular dynamics simulations examine how local contact geometry and pre-existing twin configurations affect nucleation, propagation, and thickening, providing a qualitative framework for interpreting the experimental observations within the limitations of the atomistic length and time scales. This approach establishes an experimental link between prior deformation history and the distribution of twinning onset stresses and assesses how the availability of twin embryos and mobile interfaces contributes to the observed variability. The resulting contribution is a clearer understanding of how the local microstructural state influences the first twinning-related plastic event under otherwise comparable geometric and crystallographic loading conditions.

\section{Materials and methods}
\subsection{Sample preparation}

Commercially pure Mg (99.98\%) was used to grow single crystals via the vertical Bridgman technique, as described in \cite{molodov2022effect}. Monocrystalline samples with dimensions of 14 mm $\times$ 10 mm $\times$ 6 mm were extracted from the as-grown crystal using electrical discharge machining (EDM). The crystallographic orientations were determined using Laue diffraction, with the 14 mm, 10 mm, and 6 mm dimensions aligned parallel to \hkl[0001], \hkl[11-20], and \hkl[1-100], respectively. 

Plain strain compression (PSC) in a channel-die setup was employed to pre-deform the samples and introduce \hkl{10-12} tensile twins. As illustrated in Figure \ref{fig_Experimental_Setup} (a), the samples were compressed along the \hkl[1-100] direction, resulting in extension along the \hkl[0001] direction, which aligns with the longitudinal direction (LD) of the channel-die. The \hkl[11-20] direction aligns with the transverse direction (TD) of the channel die, where lateral strains are geometrically constrained by the channel-die walls. PSC was conducted at room temperature (RT) and a strain rate of $10^{-3} \, \text{s}^{-1}$, up to a total strain of only 2\%, in order to activate twinning while minimizing the activation of competing deformation mechanisms. Following PSC, the samples were sectioned by EDM along the TD direction, such that \hkl[11-20] forms the surface normal for both the matrix and twin regions.  The exposed surfaces were prepared by standard metallographic procedures, including grinding with SiC papers and mechanical polishing using ethanol-based diamond suspensions down to 1 µm. Final surface preparation was achieved by electropolishing in AC2 electrolyte for 90 s at -20 °C and 20 V. 

\subsection{Micropillar compression}
A scanning electron microscope (SEM) with focused ion beam (FIB) (Helios NanoLab 600i DualBeam, Thermo Fisher Scientific) was used to identify regions of interest and fabricate micropillars. Initial imaging was performed using SEM, followed by electron backscatter diffraction (EBSD) to determine crystallographic orientations. As illustrated in Figure \ref{fig_Experimental_Setup} (b), for both the matrix and twin grains, the \hkl[11-20] crystal direction is aligned with the TD sample direction. The sample was rotated about the TD to align the TB vertically. This configuration ensures that the orientations of the matrix and the adjacent twin grain are symmetric with respect to the TB plane. Two arrays of micropillars were fabricated within the matrix grain and the adjacent twin grain using an annular FIB milling protocol  (Figure \ref{fig_Experimental_Setup} (c)), and are denoted as MG and TG, respectively. Milling was carried out at an accelerating voltage of 30 kV, starting with a beam current of 9.3 nA and progressively reducing the current in successive steps, with a final polishing step at 80 pA. All pillars had a diameter of 4 µm and an approximate aspect ratio of 2:1 (height to top-surface diameter). The resulting taper angle was below 2°, and is therefore not expected to cause significant stress inhomogeneities within the micropillars or affect the measured plastic response \cite{kiener2009micro}.

Micropillar compression tests were conducted at room temperature using an FT-NMT04-XYZ in-situ nanoindenter from FemtoTools integrated into a TESCAN CLARA SEM, equipped with a 10 µm diameter diamond flat punch indenter. Based on established findings \cite{liu2017experimentally,jeong2018situ,della2023micromechanical,della2023temperature}, the strain rate and micropillar dimensions were deliberately selected to favor activation of \hkl{10-12} twinning over other deformation mechanisms. All tests were performed in a displacement-controlled mode at a constant loading speed of 9 nm $\text{s}^{-1}$, which corresponds to an approximate strain rate of $10^{-3} \, \text{s}^{-1}$ considering an average pillar height of 9 µm. While mostly an engineering strain of approximately 8–12\% was targeted, several interrupted tests were intentionally manually stopped after the first plastic event, which occurred at an engineering strain of less than 2\%. The engineering strain is based on the measured displacement divided by the initial pillar height, and the engineering stress is based on the measured load divided by the pillar's upper cross section. The reported strain is therefore a nominal strain that can include contributions from contact deformation and deformation of the supporting substrate. Consequently, the displacement-based curves presented here were not used to determine the intrinsic elastic modulus. At least 10 micropillars were tested under identical conditions to ensure reproducibility. Post-mortem microstructural characterization was performed on cross-sections of selected deformed micropillars prepared via FIB lift-out. EBSD measurements were conducted at an accelerating voltage of 15 kV and a beam current of 5.5 nA, with a step size of 50 nm. Data analysis was carried out using EDAX OIM Analysis™ software.

\begin{table*}[htp!]
\centering
\caption{Schmid factors and reported CRSS values of possible deformation modes during \hkl[11-20] micropillar compression.}
\label{tab:deformation}
\footnotesize 
\begin{tabular*}{\textwidth}{@{\extracolsep{\fill}}lcccccc}
\toprule
\multirow{3}{*}{\begin{tabular}[c]{@{}l@{}}Deformation\\ Mode\end{tabular}} & 
\multirow{3}{*}{\begin{tabular}[c]{@{}c@{}}Slip/Twin\\ System\end{tabular}} & 
\multirow{3}{*}{\begin{tabular}[c]{@{}c@{}}Schmid\\ factor ($m$)\end{tabular}} & 
\multirow{3}{*}{\begin{tabular}[c]{@{}c@{}}CRSS\\ (MPa)\end{tabular}} & 
\multirow{3}{*}{\begin{tabular}[c]{@{}c@{}}CRSS/$m$\\ (MPa)\end{tabular}} & 
\multicolumn{2}{c}{After reorientation by} \\
& & & & & \multicolumn{2}{c}{tensile twinning} \\
\cmidrule{6-7}
& & & & & Schmid factor ($m$) & CRSS/$m$ (MPa) \\
\midrule
Basal $\langle a \rangle$ & \hkl(0001)\hkl[2-1-10] & 0 & 0.52 \cite{conrad1957effect} & / & 0.45 & 1.16 \\
Prismatic $\langle a \rangle$ & \hkl(01-10)\hkl[2-1-10] & 0.43 & 39 \cite{reed1957deformation} / 50 \cite{flynn1961thermally} & 91 / 116 & 0.12 & 325 / 417 \\
Pyramidal II $\langle c+a \rangle$ & \hkl(11-22)\hkl[11-2-3] & 0.45 & 80 \cite{ando199211} & 178 & 0.19 & 421 \\
Tensile twin & \hkl(10-12)\hkl[-1011] & 0.37 & 12 \cite{kelley1968plane} & 32 & 0.23 & 52 \\
\bottomrule
\label{tableCRSS}
\end{tabular*}
\end{table*}

\begin{figure*}[h!]
\centering
\includegraphics[width=0.8\textwidth]{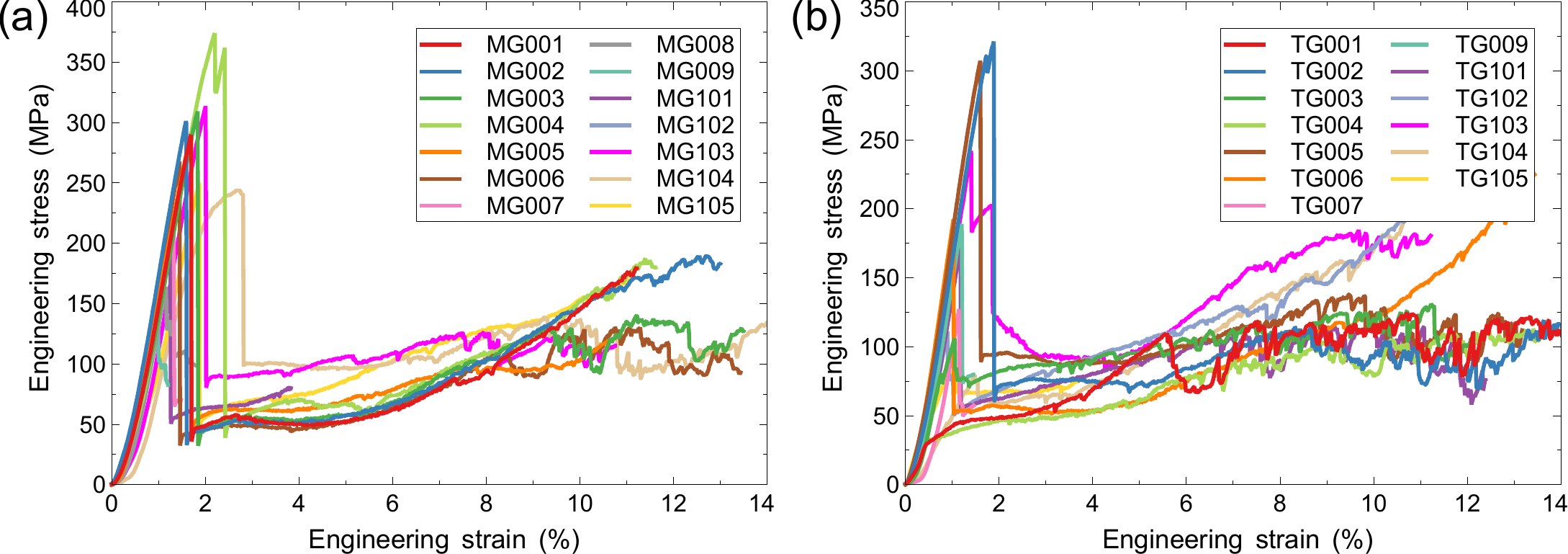}
\caption{Engineering stress-strain curves of \hkl[11-20] micropillar compression tests for (a) matrix grain (MG) and (b) twin grain (TG).}
\label{fig_Stressstrain}
\end{figure*}

\subsection{Atomistic simulation}

Molecular dynamics (MD) simulations were conducted using the Large-scale Atomic/Molecular Massively Parallel Simulator (LAMMPS) \cite{thompson2022lammps} to resolve local twinning events that cannot be separated temporally in the micropillar experiments and to test how contact geometry and pre-existing defects alter the activation sequence of twinning events within each specified simulation setup. Interatomic interactions in Mg were described using the modified embedded atom method (MEAM) potential developed by Kim et al. \cite{kim2015modified}. Our previous systematic assessment of various semi-empirical potentials for Mg \cite{wang2024defects} demonstrated that this potential aligns well with density functional theory (DFT) and experimental measurements in terms of lattice constants, elastic constants, stacking fault energies, TB energies, and dislocation properties. Furthermore, complementary simulations were performed using an alternative MEAM potential \cite{lee2024atomistic}, developed for atomistic investigation of deformation twinning in single-crystalline Mg under compression, to test whether the main mechanistic observations are potential-dependent. Atomsk \cite{hirel2015atomsk} was utilized to construct nanopillar samples with a diameter of 30 nm and a height of 60 nm, matching the experimental aspect ratio and comprising approximately 1.8 million atoms. The crystallographic orientations were defined such that the x-, y-, and z-axes correspond to the \hkl[0001], \hkl[1-100], and \hkl[11-20] directions, respectively. Non-periodic boundary conditions were applied along all three directions. All structures were first energy-minimized using the fast inertial relaxation engine (FIRE) algorithm \cite{bitzek2006structural,guenole2020assessment} with a force tolerance of 10$^{-8}$ eV/\AA. Following equilibration at 10 K in the NVT ensemble with the Nosé-Hoover thermostat \cite{hoover1985canonical}, uniaxial compression was applied along the z-axis using virtual indenters at 10 K. A timestep of 1 fs and a strain rate of $10^{8}$ $\text{s}^{-1}$ were employed. Post-processing and visualization were performed using OVITO \cite{stukowski2009visualization}. Structural analysis was performed using polyhedral template matching (PTM) \cite{larsen2016robust} and common neighbor analysis (CNA) \cite{honeycutt1987molecular}.

\begin{figure*}[h!]
\centering
\includegraphics[width=\textwidth]{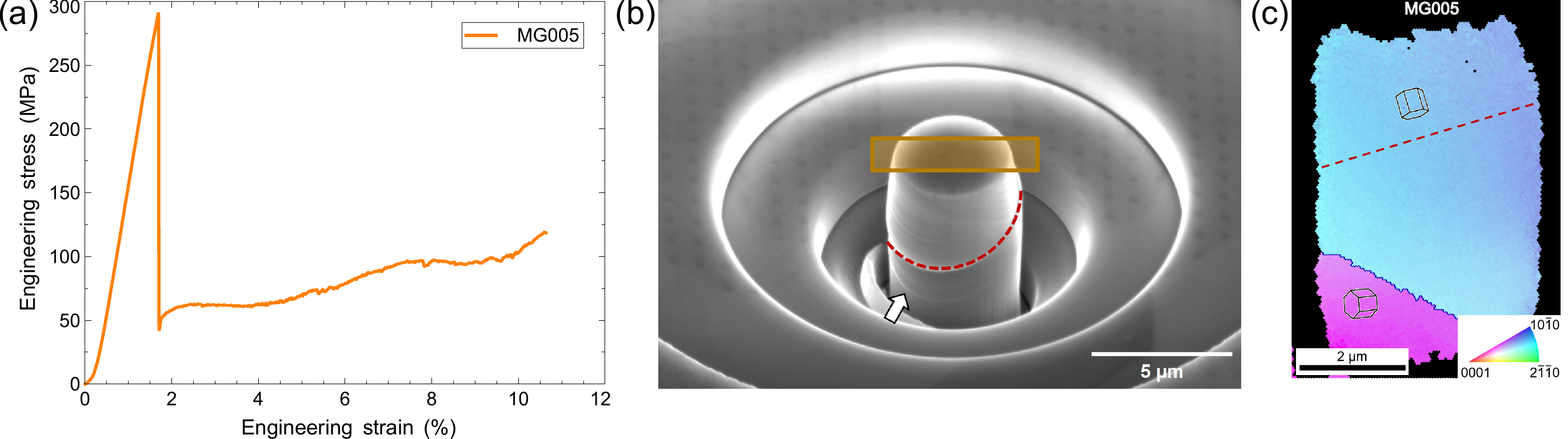}
\caption{Representative example of the mechanical and microstructural characterization of the deformed pillar MG005. (a) Engineering stress-strain curve. (b) SE image after compression, captured at a $45^{\circ}$ tilt angle. (c)  IPF-Z map of the cross-section of the pillar.}
\label{fig.MG005_all}
\end{figure*}

\section{Results}

\subsection{Expected deformation mechanisms in \hkl[11-20]-oriented pillars }
The possible deformation modes and corresponding Schmid factors ($m$) of single-crystal Mg micropillars deformed in \hkl[11-20] compression are summarized in Table \ref{tableCRSS}. The Schmid factor for basal \hkl<a> slip is zero. Despite its low reported CRSS of only 0.52 MPa \cite{conrad1957effect}, basal slip is considered to be effectively suppressed under \hkl[11-20] compression. In contrast, prismatic \hkl<a> slip and pyramidal II \hkl<c+a> slip systems are geometrically well-oriented ($m > 0.43$), but the high stresses usually required for their activation, as indicated by published CRSS values \cite{reed1957deformation,flynn1961thermally,ando199211}, suggest they remain inactive at the onset of plastic deformation. Based on the deformation response,  \hkl{10-12} tensile twinning emerges as the most likely initial deformation mode. Notably, four variants of tensile twins share an identical high Schmid factor of \(m = 0.37\). Once one or several of these twins form within the micropillars, the local lattice reorients such that the compression axis in the twinned region becomes \hkl[11-23] \cite{kim2011small}, for which the Schmid factor for basal slip is maximal. Basal slip is therefore anticipated to dominate subsequent deformation.
\subsection{Deformation response and microstructure evolution}
The engineering stress-strain curves obtained from the micropillar compression tests are shown in Figure \ref{fig_Stressstrain}. Most micropillars from both the matrix (Figure \ref{fig_Stressstrain} (a)) and the adjacent twin-grain (Figure \ref{fig_Stressstrain} (b)) exhibit a similar multi-stage deformation behavior. Following an initial elastic regime, a pronounced stress drop occurs at an engineering strain of approximately 2\%, succeeded by a flow-stress plateau. As strain increases, a relatively stable plastic flow at stress levels between 50 MPa and 100 MPa is observed, indicating the activation of softer deformation modes compared to the first plastic event. Focusing on this first event, the stress associated with the sharp stress drop is around 300 MPa for most micropillars in the matrix-grain. In contrast, only two of the fourteen micropillars in the twin-grain reach similar values, whereas the remaining pillars display comparatively lower peak stresses between 100 MPa or 180 MPa. This reflects differences in activation stresses and a more heterogeneous mechanical response in the twin-grain. Notably, pillars TG001 and TG004 show the lowest flow stresses (approximately 30 MPa) and do not display a discernible stress drop. The deformation behavior of parent-grain and twin-grain pillars will be discussed individually in the following sections.

\subsubsection{Parent-grain pillars}
To establish a correlation between the mechanical response of the tested micropillars and their corresponding deformation microstructures, a detailed post-mortem EBSD analysis was performed on pillar cross-sections. This analysis comprised orientation maps and kernel average misorientation
(KAM) mapping. Pillar MG005 is used in Figure \ref{fig.MG005_all} as a representative example. Its stress-strain curve shows a characteristic sharp stress drop followed by a flow stress plateau with slight strain hardening (accounting for geometric effects) and several minor stress drops (Figure \ref{fig.MG005_all} (a)). A secondary electron image after $\sim$11\% compressive strain is shown in Figure \ref{fig.MG005_all} (b). Distinct slip steps are visible on the pillar surface and are highlighted by the red dashed line and the white arrow. To investigate the underlying deformation structure, EBSD measurements were performed on cross-sections prepared by FIB lift-out. The central region of the micropillars (yellow rectangle in Figure \ref{fig.MG005_all} (b)) was selected for the analysis. The inverse pole figure (IPF-Z) orientation map of the cross-section of pillar MG005, shown in Figure \ref{fig.MG005_all} (c) reveals a pillar-spanning twin formed during compression. The twin is shown in blue, whereas the remaining matrix is magenta, reflecting the sample rotation about the TD.  The TB is marked in dark blue, and the measured lattice orientations are consistent with the characteristic matrix/twin misorientation of 86°\hkl<11-20> expected for a \hkl{10-12} tensile twin. Local orientation gradients within the twin, visible as a color variation in the IPF map, indicate progressive lattice rotation during deformation, leading to local variations in Schmid factor for subsequently activated slip. The dashed red line marks the basal-plane trace within the twin and aligns with the slip steps observed on the pillar surface in Figure \ref{fig.MG005_all} (b). Based on these observations, the initial sharp stress drop is attributed to the onset of tensile twinning, whereas the subsequent smaller stress fluctuations are ascribed to basal \hkl<a> slip within the reoriented twin volume. 

\begin{figure}[h!]
\centering
\includegraphics[width=0.5\textwidth]{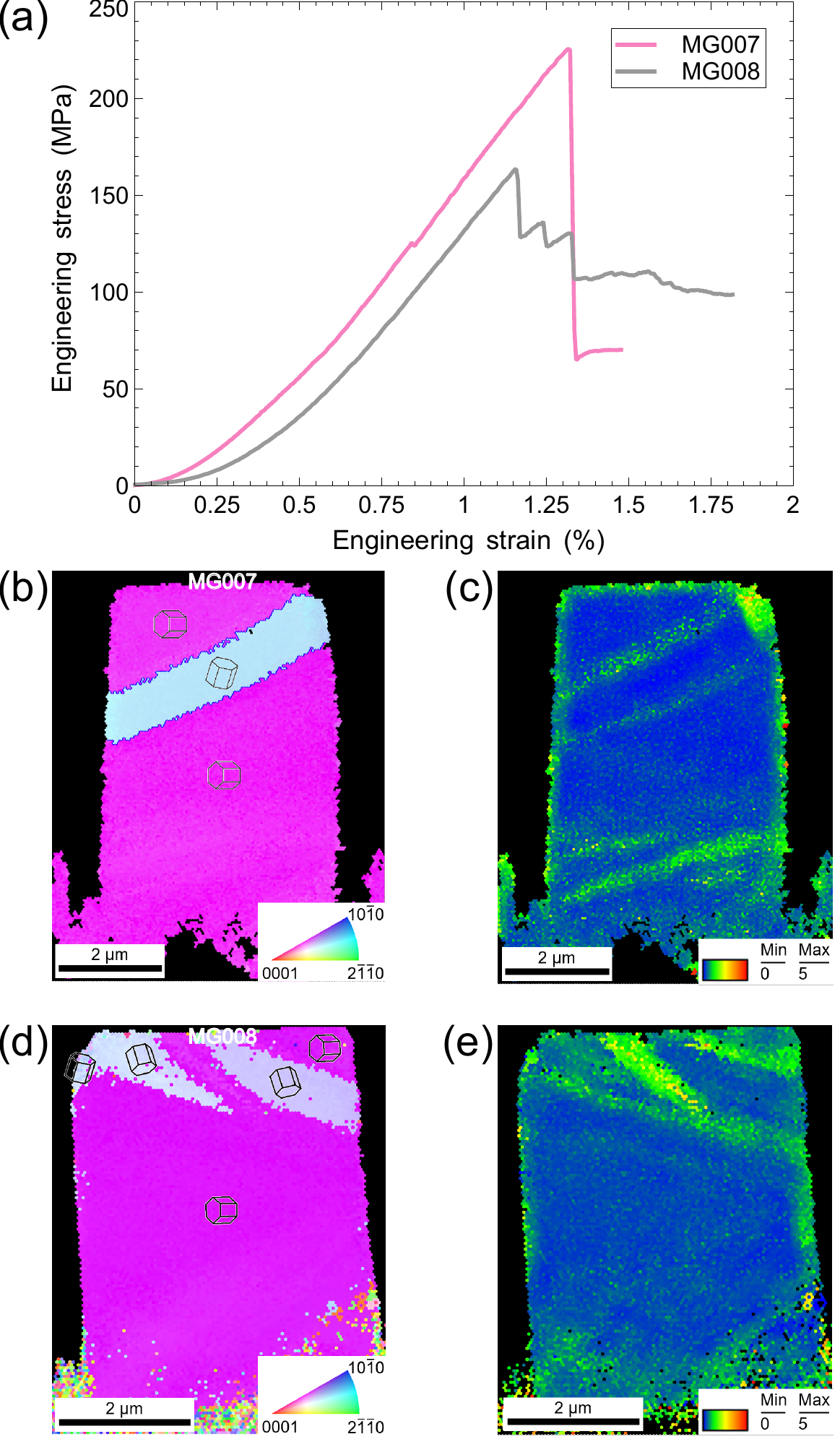}
\caption{Interrupted microcompression tests of pillars MG007 and MG008 and their corresponding post-mortem microstructural characterization. (a) Engineering stress-strain curves for pillar MG007 and MG008. (b, d) IPF-Z maps and (c, e) corresponding KAM maps of pillars MG007 and MG008, respectively. The KAM maps were constructed by averaging the misorientation of a point with its 3rd nearest neighbors.}
\label{fig.MG007008}
\end{figure}

To corroborate that the initial load drop at the onset of the first plastic event is associated with the formation of a \hkl{10-12} tensile twin, interrupted micropillar compression tests were performed to different strain levels. In these experiments, loading was manually stopped immediately after the pronounced stress drop had been observed, in order to isolate this specific event. The engineering stress-strain curves for the interrupted tests MG007 and MG008 are shown in Figure \ref{fig.MG007008} (a). For MG007, a sharp stress drop occurs at a strain of 1.3\%, corresponding to a peak stress of 225 MPa. Compression was terminated after a brief strain burst reaching a total strain of 1.5\%. The IPF-Z map of the MG007 cross-section in Figure \ref{fig.MG007008} (b) reveals a dominant single twin band that has traversed a significant portion of the pillar. The KAM map in Figure \ref{fig.MG007008} (c) provides further insight into the plastic strain distribution, showing elevated KAM values (up to 2°) concentrated along the TBs. Maximum KAM values (5°) at the upper right of the pillar suggest that the twin nucleated at the pillar edge, which serves as a stress concentration site, and subsequently propagated rapidly across the full pillar width before thickening occurred through TB migration at a nearly constant flow stress of $\sim$65 MPa. 

Compared with MG007, pillar MG008 exhibits three pronounced stress drops on its stress-strain curve. Consistently, three distinct twins are observed near the top of the pillar in the IPF-Z map of the MG008 cross-section (Figure \ref{fig.MG007008} (d)). Notably, the unit cell orientations indicate that two of these twins belong to the same twin variant, as four twin variants share the same Schmid factor under \hkl[11-20]-compression. These observations support the interpretation that each discrete stress drop in the stress–strain curve is associated with an individual twinning event. The KAM structure in Figure \ref{fig.MG007008} (e) shows elevated local misorientations in the vicinity of the TBs and within the matrix region confined between two twins, implying an increased accumulation of geometrically necessary dislocations (GND) in these regions. It should be noted that the variation in aspect ratio observed in the pillar EBSD maps arises from non‑diametric cross‑sectioning of the cylindrical pillars and from cropping to retain only regions with high indexing confidence. This processing does not affect the validity of the crystallographic interpretation.

\begin{figure}[h!]
\centering
\includegraphics[width=0.5\textwidth]{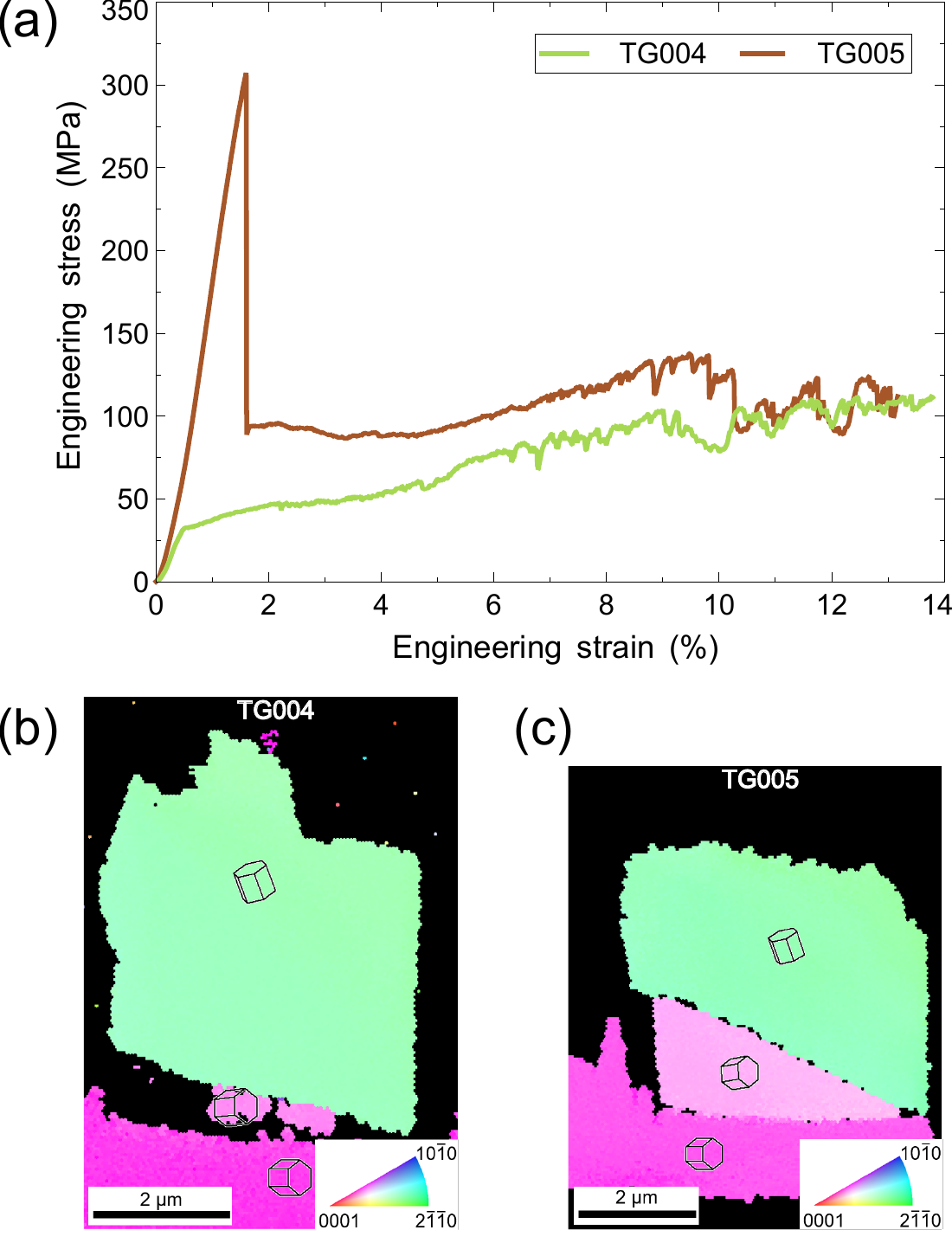}
\caption{Microcompression tests of pillars TG004 and TG005 and their corresponding post-mortem microstructural characterization. (a) Engineering stress-strain curves. (b, c) Cross-sectional IPF-Z maps of (b) TG004 and (c) TG005, with the twinned region shown in green and the parent matrix in pink. The magenta region at the bottom corresponds to the underlying bulk material beneath the pillar base.}
\label{fig.TG004005}
\end{figure}
 \subsubsection{Twin-grain pillars}
Figure \ref{fig.TG004005} (a) compares the engineering stress-strain curves of two pillars in the twin-grain region and reveals markedly distinct mechanical responses. Pillar TG005 demonstrates a similar response to the parent-grain pillars, characterized by an abrupt stress drop at $\sim$300 MPa, at the onset of yield, whereas pillar TG004 yielded at a much lower stress (32 MPa) without a discernible stress drop. Although site-specific FIB lift-out of the heavily deformed pillars introduced local artifacts and partial material loss, the IPF maps nevertheless unambiguously identify deformation twinning as the dominant deformation mode in both TG004 and TG005 (Figure \ref{fig.TG004005} (b, c)). Moreover, the lattice orientations indicate activation of the same twin variant in both pillars. Given that deformation twinning proceeds via nucleation, longitudinal propagation, and lateral thickening \cite{christian1995deformation}, the large disparity in the apparent yield stress between TG004 and TG005 suggests that different stages of the twinning process controlled the observed response. 

\begin{figure}[h!]
\centering
\includegraphics[width=0.5\textwidth]{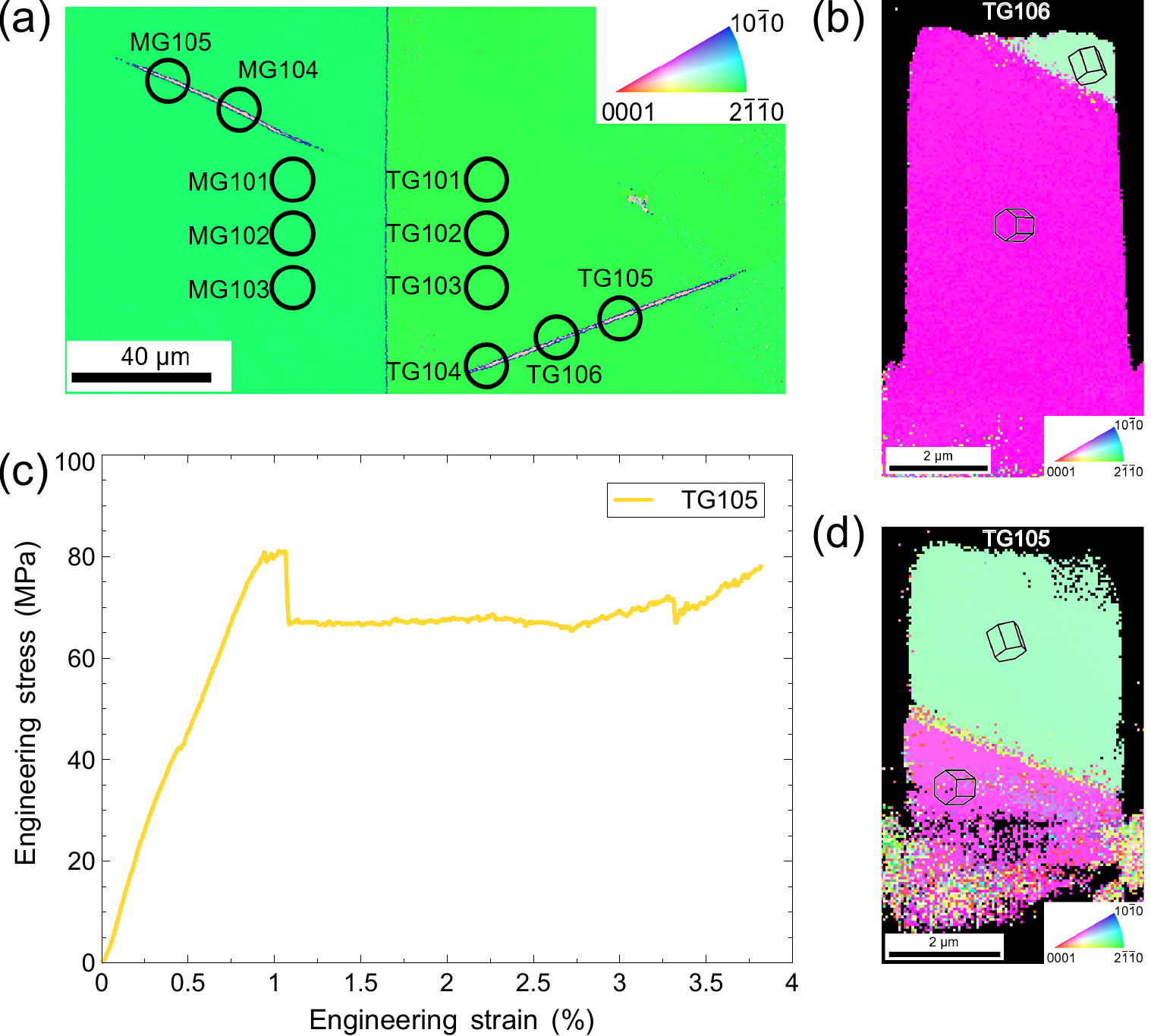}
\caption{Microstructural characterization of pillars TG105 and TG106 with a pre-existing secondary twin. (a) IPF-TD map of the region of interest showing the matrix-grain (left) and the twin-grain (right). Labeled black circles mark the micropillar locations. (b) Cross-sectional IPF-Z map of the undeformed TG106. (c) Engineering stress-strain curve of pillar TG105. (d) Cross-sectional IPF-Z map of pillar TG105 after $\sim$4\% deformation.}
\label{fig.TG105106}
\end{figure}

This interpretation was examined using two site-specifically fabricated pillars, TG105 and TG106, containing pre-existing secondary twins within the primary twin-grain (Figure \ref{fig.TG105106} (a)). TG106 was kept undeformed as a microstructural reference, while TG105 was compressed to a strain of approximately 4\%. The cross-sectional IPF-Z map of TG106 (Figure \ref{fig.TG105106} (b)) confirms the presence of a secondary twin-grain prior to deformation, located at the upper-right end of the pillar contacting the punch. The mechanical response of TG105 reveals a short yield plateau at a relatively low stress of $\sim$81 MPa, followed by a discrete stress drop at $\sim$1.1\% strain and subsequent plastic flow at a constant stress of $\sim$68 MPa (Figure \ref{fig.TG105106} (c)). Corresponding post-mortem EBSD analysis (Figure \ref{fig.TG105106} (d)) shows that the initial secondary TB has migrated downward from the pillar top, providing direct evidence for lateral twin thickening via TB migration during deformation. Notably, the stress associated with this thickening process ($\sim$68 MPa) agrees well with the post-drop flow-stress levels observed in the majority of other pillars ($\leq$ 100 MPa). This indicates that a pre-existing twin can accommodate plastic strain by boundary migration (twin thickening) without a new high-stress nucleation event.

\subsection{MD simulations of local twinning events}
To complement the experimental observations and qualitatively probe the underlying atomic-scale mechanisms, MD simulations were performed to examine the distinct stages of twin nucleation, longitudinal propagation, and lateral thickening. Tailored atomistic setups were designed to decouple these processes conceptually and to clarify how local defect geometry alters the onset of twinning. Two complementary visualization methodologies were implemented based on the PTM algorithm. The first approach utilized PTM to characterize defect structures. Atoms identified as HCP were filtered out, and the remaining non-HCP atoms were rendered according to their local structure (FCC: green; BCC: blue; “Other”: white). This method clearly delineates the TB position, given that interfacial atoms are typically classified as “Other". In the second approach, atoms were colored by their local misorientation of the surrounding lattice relative to the initial matrix, as calculated via PTM. Following the characteristic matrix/twin crystallographic relationship of $\sim$90° lattice reorientation for \hkl{10-12} twins, matrix atoms were colored magenta and reoriented twin atoms blue, to match the coloration in the experiments. This representation provides an intuitive visualization of twin morphology and enables direct tracking of the evolving twin volume fraction during the simulation.

\subsubsection{Twin nucleation}

\begin{figure*}[h!]
\centering
\includegraphics[width=0.8\textwidth]{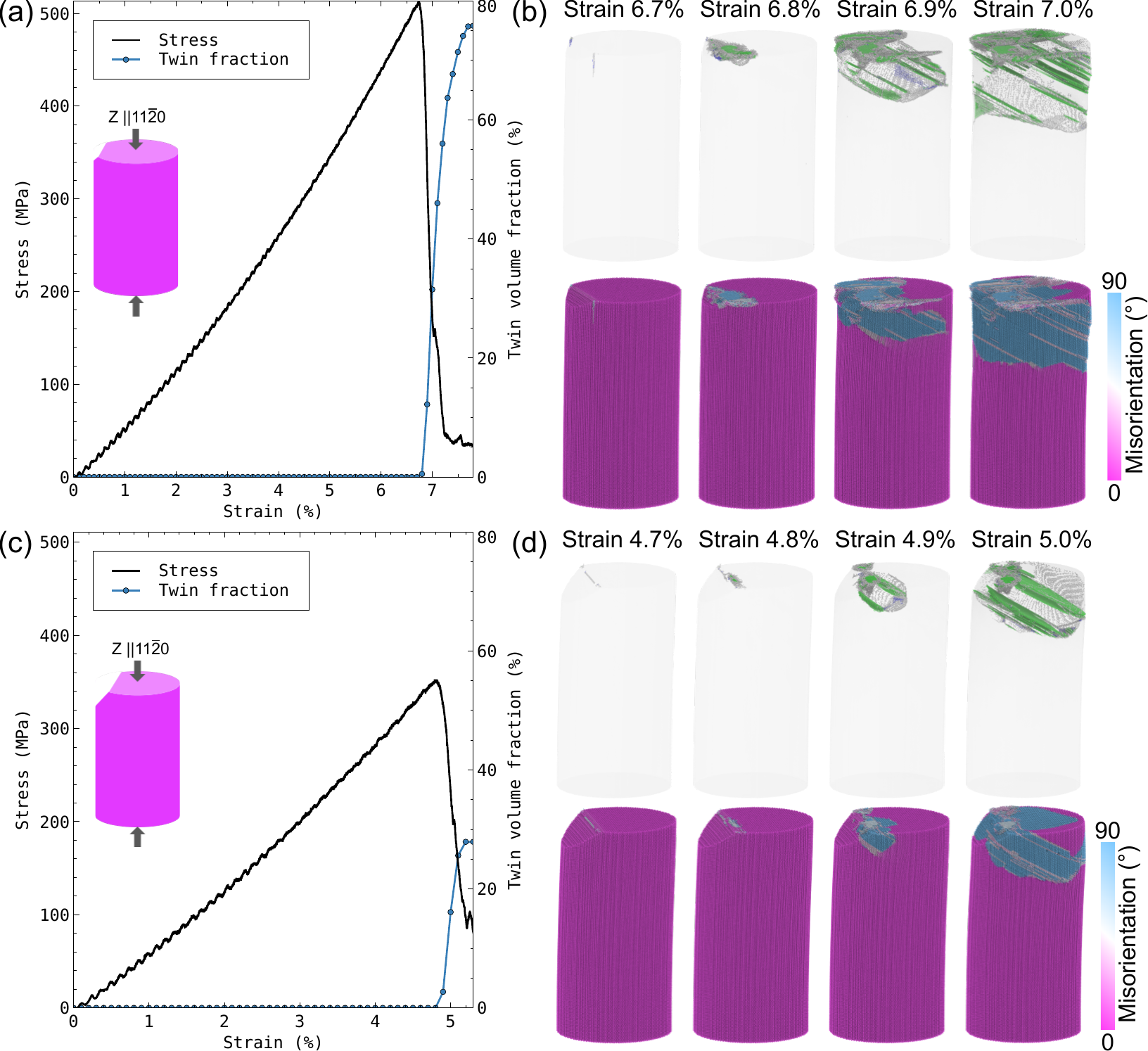}
\caption{Comparative MD simulations of twin nucleation. (a, c) Mechanical response and twin volume fraction evolution for  Setup 1 (a) and Setup 2 (c), including pillar initial configurations. (b, d) Corresponding snapshots of the nucleation process. In each snapshot, the top row highlights non-HCP defect atoms identified by PTM together with surface meshes, and the bottom row show atomic misorientation relative to the matrix.}
\label{fig.nucleation}
\end{figure*}

Two MD setups were constructed to investigate the twin nucleation process under different contact surface conditions, as shown in Figure \ref{fig.nucleation}. Motivated by the experimental observation of twin formation near stress concentrations (Figure \ref{fig.MG007008}), the initial pillar structure of Setup 1 (Figure \ref{fig.nucleation} (a)) incorporates a truncated top corner. This geometric design introduces a localized stress field at the indenter-pillar contact to promote twin nucleation \cite{jiang2022visualization}. Uniaxial compression was applied using virtual indenters along the pillar z-axis, aligned with the \hkl[11-20] direction. 

In Setup 1, the nominal axial stress reaches a pre-event peak of 511 MPa immediately before the first stable twin embryo appears. Consistent with the experimental observations, a prominent stress drop occurs at an imposed strain of 6.8\%, accompanied by a rapid increase in twin volume fraction from 0\% to 75.7\% over a narrow strain interval between 6.7\% and 7.7\% strain. The snapshots in Figure \ref{fig.nucleation} (b) covering the interval from 6.7\% to 7.0\% strain, show embryo formation at the truncated corner followed by rapid expansion across the pillar. The abrupt increase in twin volume fraction therefore reflects nucleation followed by extensive twin propagation, with the accompanying stress drop indicating plastic relaxation (cf. Section ~\ref{sec:propagation}).

In Setup 2, the truncated corner is enlarged to represent an altered contact surface condition (Figure \ref{fig.nucleation} (c,d)). Twinning again governs the initial plastic response. Two twin embryos form at distinct locations near the truncated corner, become visible at an imposed strain of 4.9\% and coalesce into a single twin region by 5.0\% strain. The pre-event peak stress is 352 MPa, compared with 511 MPa in Setup 1. This difference illustrates the sensitivity of the observed twinning onset stress to contact geometry under the simulated loading conditions. 

In both setups, defect-structure snapshots reveal multiple basal stacking faults within the reoriented twinned region, consistent with basal \hkl<a> slip activity. The sequence of localized twin formation, rapid growth and stress relaxation, together with slip activity within the twin, provides a qualitative correspondence with the experimental observations. 

\subsubsection{Twin propagation}
\label{sec:propagation}

\begin{figure*}[h!]
\centering
\includegraphics[width=0.8\textwidth]{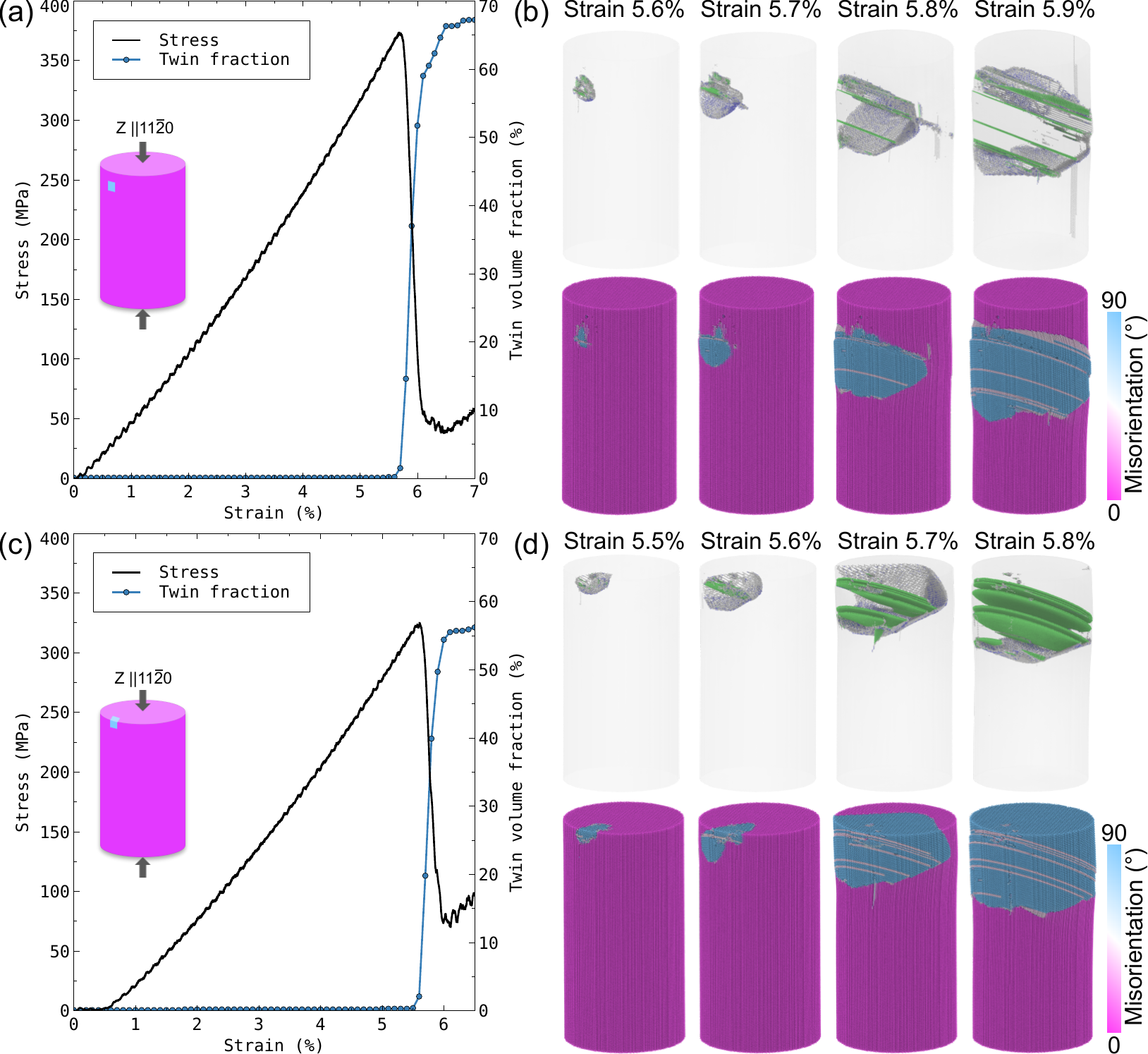}
\caption{Comparative MD simulations of twin propagation. (a, c) Mechanical response and twin volume fraction evolution for Setup 1 (a) and Setup 2 (c), including initial pillar configuration. (b, d) Corresponding snapshots of the propagation process through the pillar. In each snapshot, the top row highlights non-HCP defect atoms identified by PTM together with surface meshes, and the bottom row show atomic misorientation relative to the matrix.}
\label{fig.propagation}
\end{figure*}

To examine twin propagation from a pre-existing embryo, a twin embryo was introduced into the pillar using the Eshelby method \cite{xu2013importance}. Two initial configurations were considered: An embryo located at the pillar side surface in Setup 1  (Figure \ref{fig.propagation} (a)) and at the top surface in Setup 2 (Figure \ref{fig.propagation} (c)). Upon loading, the applied stress rises until the existing embryo begins to expand, reaching 371 MPa in Setup 1 and 324 MPa in Setup 2. These values characterize the setup-specific stress peaks preceding embryo expansion and should not be interpreted as the stresses required to sustain propagation. 

In Setup 1, the rapid twin advances across the accessible portion of the pillar within a narrow imposed-strain interval (5.6\% to 5.9\%), accompanied by a pronounced stress drop (Figure \ref{fig.propagation} (b)). This response indicates rapid plastic relaxation once embryo expansion begins. During this interval, thickening along the pillar axis is comparatively limited. Expansion toward the upper pillar region is locally impeded by dislocation obstacles that restrict TB migration in that direction. The activation of basal \hkl<a> slip within the reoriented twinned region is also evident in this setup. At a strain of 5.8\%, the snapshot of the defect structure shows transmission of a basal \hkl<a> dislocation across the TB, which continues to slip as a prismatic \hkl<a> dislocation within the matrix. 

In Setup 2, the twin originating at the top surface has traversed the upper pillar region by an imposed strain of 5.8\% (Figure \ref{fig.propagation} (d)). Subsequent growth proceeds predominantly through lateral thickening. Together, the two configurations illustrate how the initial embryo location and local obstacles influence the spatial evolution of twin growth. 

\subsubsection{Twin thickening}

\begin{figure*}[h!]
\centering
\includegraphics[width=0.8\textwidth]{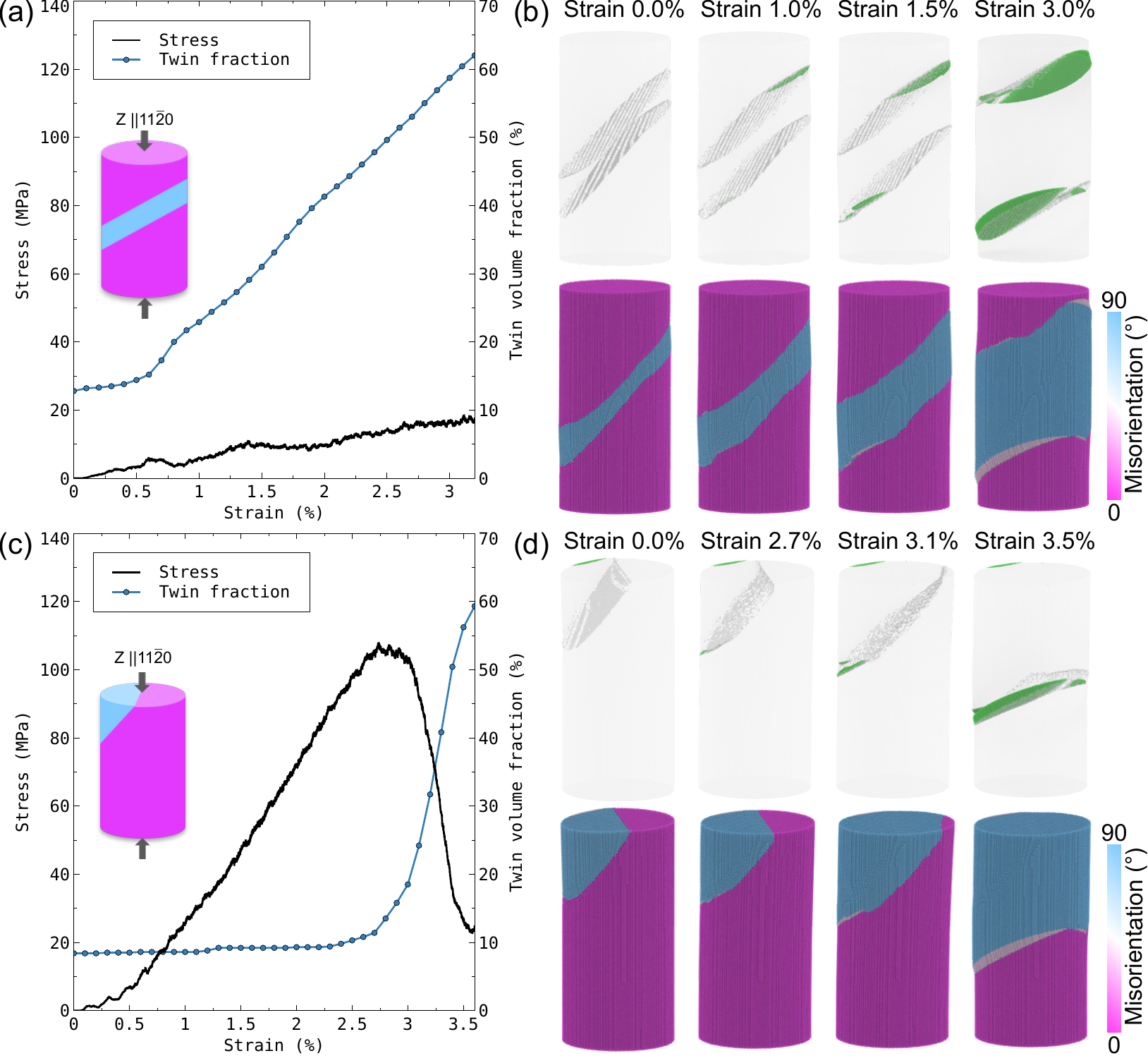}
\caption{Comparative MD simulations of twin thickening. (a, c) Mechanical response and twin volume fraction change for (a) Setup 1 and (c) Setup 2, including initial configuration sketches. (b, d) Corresponding snapshots of the thickening process. For both setups, the top row highlights defect atoms using PTM and surface meshes, while the bottom row color-codes atoms based on their misorientation relative to the matrix.}
\label{fig.thickening}
\end{figure*}

To investigate twin thickening through migration of pre-existing interfaces, a well-defined twin slab was first constructed within a fully periodic simulation box along all three axes, and then converted into a nanopillar geometry similar to those used in the previous setups. Setup 1 is illustrated in Figure \ref{fig.thickening} (a), where the twin slab is placed at the pillar center. After relaxation, the twin slab is bounded by TBs containing disconnections, as shown in Figure \ref{fig.thickening} (b). During compression, thickening begins at an imposed strain of 0.6\% and a very low axial stress of 6 MPa. The stress-strain response exhibits continuous plastic flow without pronounced stress drops. The progressive increase in twin volume fraction is accompanied by migration of the bounding interfaces normal to the twinning plane, predominantly along the pillar axis. The snapshots in Figure \ref{fig.thickening} (b) illustrate this evolution with increasing imposed strain.

In Setup 2, the twin slab is positioned at the top corner of the pillar (Figure \ref{fig.thickening} (c)) providing a simplified representation of the near-surface twin configuration observed in the experimental micropillars TG105 and TG106 shown in Figure \ref{fig.TG105106}. Under compression, twin growth proceeds in two stages. Between 2.7\% and 3.1\% strain, the migrating boundary approaches the pillar edge while the twin volume fraction increases gradually. The nominal axial stress associated with the onset of thickening in this configuration is 106 MPa. Once the boundary intersects the edge, further growth shifts exclusively to lateral thickening, accompanied by a steeper increase in twin volume fraction with imposed strain.  

The differences between the two setups illustrate the sensitivity of the simulated thickening response to the initial twin configuration and the evolution of boundary–surface interactions. The reported stress levels and changes in twin volume fraction characterize these specific nanopillar configurations and loading conditions. 

\section{Discussion}

\subsection{Interfacial processes during \hkl{10-12} tensile twinning}

The development and spatial evolution of \hkl{10-12} twins in Mg can be described in terms of embryo formation, propagation through the parent crystal and lateral thickening via migration of the bounding interfaces. Understanding these stages requires consideration of the different interface structures involved. Migration of an established coherent twin boundary (CTB) is commonly described through the motion of interfacial disconnections, which couples boundary migration to shear displacement and local atomic shuffling required to transform the parent lattice into the twin lattice \cite{serra1996new, wang20091, pond2016atomic}. A propagating embryo, however, can be bounded by a combination of coherent segments and basal–prismatic (BP/PB) facets, whose migration involves distinct interfacial configurations and defect processes \cite{xu2013importance, liu2019three}. Experimental and atomistic investigations have also identified shuffling-mediated basal–prismatic transformations during nucleation and early embryo growth under localized loading conditions \cite{jiang2022visualization}. These findings emphasize the need to distinguish the formation and expansion of a small embryo from the migration of an established coherent interface.

In this study, we examine this distinction by comparing the structural evolution of a localized twin embryo with the thickening of a pre-existing twin slab containing interfacial disconnections. Atomistic configurations and relative shear-strain maps are used to relate the observed growth to the evolving boundary morphology and the motion of identifiable interface defects. This comparison provides a basis for assessing how the initial twin configuration influences the subsequent plastic response. 

Under the simulated loading conditions, twin formation and expansion from localized embryos produce abrupt increases in twin volume fraction accompanied by pronounced stress relaxation, whereas the pre-existing twin slab thickens more gradually with imposed strain. The pronounced slowdown during thickening indicates that lateral TB migration becomes rate-limiting and does not represent a straightforward continuation of the same elementary process. The observed responses are therefore interpreted together with the atomistic configurations to examine the interfacial processes involved in each stage. 

\begin{figure*}[h!]
\centering
\includegraphics[width=\textwidth]{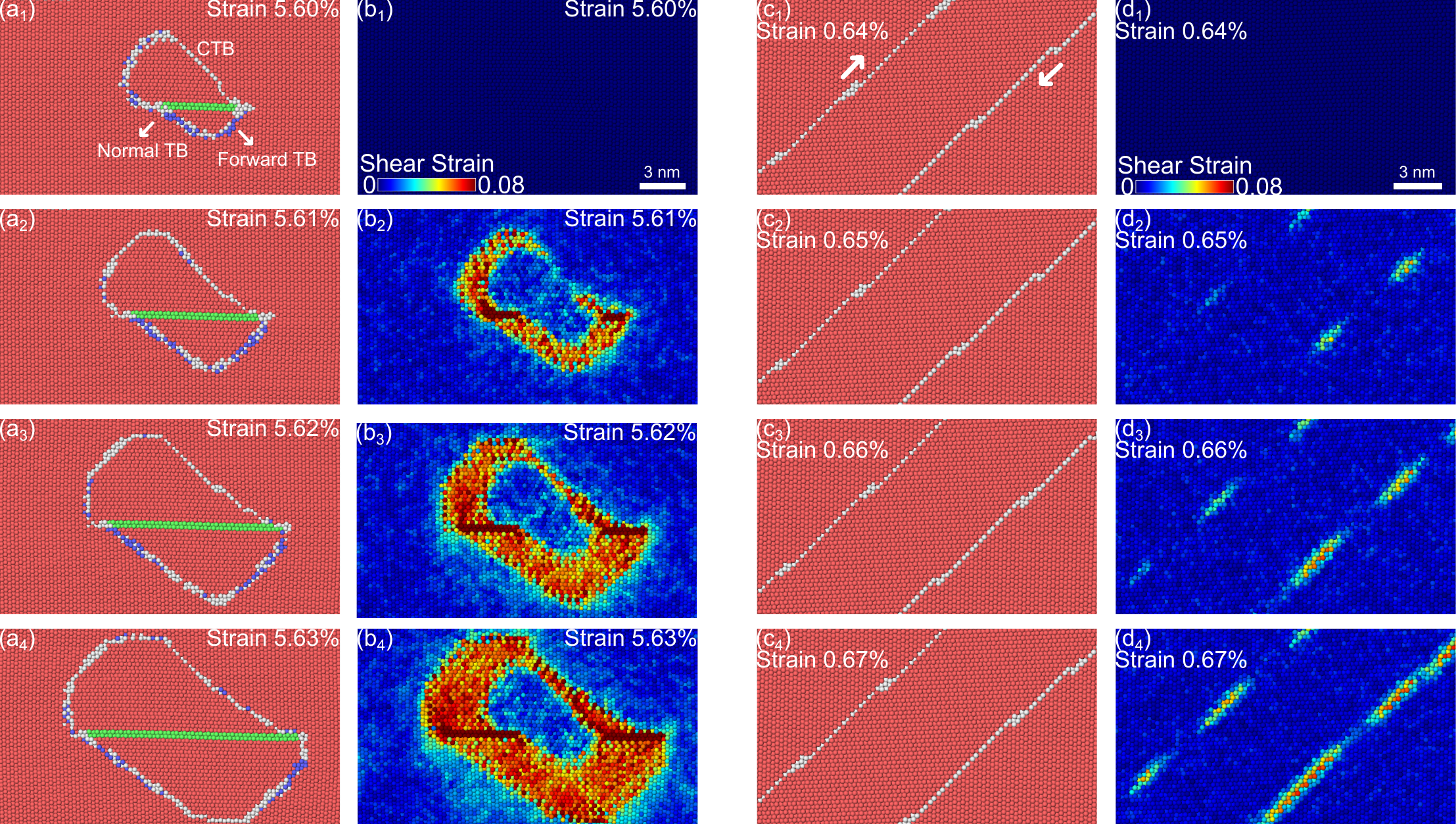}
\caption{Atomistic configurations and corresponding strain of the $\kappa_{1}$ plane during twin propagation and thickening. (a, b) Sequence for Propagation Setup 2 and (c, d) for Thickening Setup 1. Snapshots (a$_{1}$–a$_{4}$) and (c$_{1}$–c$_{4}$) are color-coded by the CNA method to identify the local lattice structure. Snapshots (b$_{1}$–b$_{4}$) and (d$_{1}$–d$_{4}$) display the corresponding relative shear strain distributions, where the initial frames (b$_{1}$ and d$_{1}$) serve as the reference configurations.}
\label{fig.crosssetion}
\end{figure*}

Figure \ref{fig.crosssetion} (a$_{1}$–a$_{4}$) shows the evolution of a twin embryo in \textit{Propagation Setup 2} (Figure \ref{fig.propagation} (d)), viewed along the normal to the $\kappa_{1}$ twinning plane. The sequence in the snapshots covers a small increment of 0.01 \%, starting from a strain of 5.6 \%. Following the terminology used to describe three-dimensional twin morphologies \cite{liu2019three}, the boundary segments are identified as \textit{normal} and \textit{forward} TB segments \cite{liu2019three}, as indicated in Figure \ref{fig.crosssetion} (a$_{1}$). The forward segments consist of \hkl{0001} basal $\parallel$ \hkl{10-10} prismatic (BP) and PB facets, whereas the normal segments comprise CTB facets and BP/PB steps. Note that a basal stacking fault (Green atoms in Figure \ref{fig.crosssetion} (a$_{1}$–a$_{4}$)) spontaneously formed inside the embryo upon relaxation does not influence the mechanism. 

During the evolution from a$_{1}$ to a$_{4}$, the forward boundary segments advance farther than the normal segments, producing directional differences in embryo growth within $\Delta\varepsilon$ = 0.03\%. This ultrafast propagation is captured by the corresponding relative shear-strain maps (Figure \ref{fig.crosssetion} (b$_{1}$–b$_{4}$)), referenced to the initial frame. These maps reveal accumulated deformation in the matrix regions swept by the advancing interfaces as these regions reorient into the twin lattice. The expanding strain distribution and evolving boundary morphology indicate spatially heterogeneous growth involving several interface segments. The concomitant interfacial roughening, together with the localized deformation at the advancing front, suggests that interface migration during propagation involves local atomic shuffling and short-range rearrangements. This interpretation is consistent with that proposed for rumpled TBs observed in dynamically deformed Mg \cite{xie2021twin}. During this process, the forward TBs advance faster than the normal TBs, in qualitative agreement with the growth anisotropy inferred from EBSD-based twin shape statistics in Mg \cite{liu2019three}. The preferential advance of the forward segments indicates that propagation is kinetically favored over thickening by normal CTB migration. Once propagation is constrained by the pillar geometry, continued twin growth proceeds predominantly through thickening by migration of the bounding interfaces.

Figure \ref{fig.crosssetion} (c$_{1}$–c$_{4}$) depicts the thickening process in \textit{Thickening Setup 1} (Figure \ref{fig.thickening} (b)) between 0.64 \% and 0.67\% strain. The initial twin slab is bounded by TBs containing interfacial disconnections. Over a strain increment of $\Delta\varepsilon$ = 0.01\%,  the disconnections glide along the boundaries, as indicated by the arrows in Figure \ref{fig.crosssetion} c$_{1}$, and their passage from c$_{1}$ to c$_{4}$ advances the interfaces normal to the twinning plane. The corresponding relative shear-strain distributions in Figure \ref{fig.crosssetion} (d$_{1}$–d$_{4}$) show deformation associated with the paths of the migrating disconnections. These observations support thickening through the motion of pre-existing interfacial defects. Within this description, the shear displacement carried by disconnection motion is coupled to the local atomic shuffles required to establish the twin lattice behind the migrating interface. The observed association between disconnection motion and boundary migration is consistent with experimental microscopy observations in Mg  \cite{xie2021twin}. 

Taken together, the configurations in Figure \ref{fig.crosssetion} reveal different spatial patterns of interface motion during embryo expansion and slab thickening. Expansion of the localized embryo involves the evolution of a faceted front, whereas migration of the extended TBs can be traced to the motion of identifiable disconnections already present before loading. The significance for the present study lies in these different initial structural conditions: the availability of an embryo or a boundary containing mobile defects changes the processes accessible during loading. This provides a qualitative basis for examining how the local microstructural state influences the first twinning-related plastic event and contributes to the experimental variability discussed in the following subsection.

\subsection{Stochastic twinning under confinement}

\begin{figure}[h!]
\centering
\includegraphics[width=0.5\textwidth]{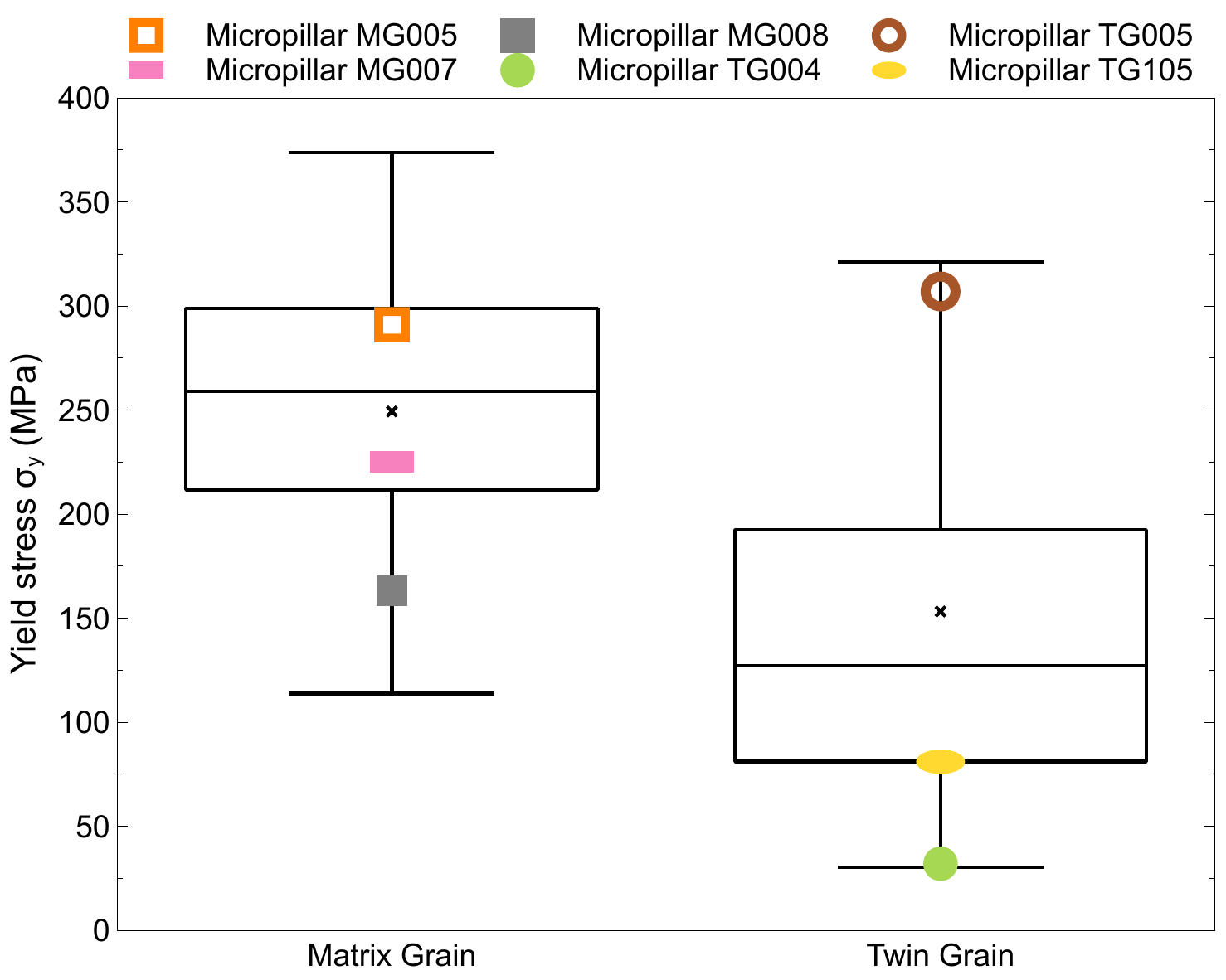}
\caption{Statistical comparison of yield stresses associated with twinning from micropillar experiments. Box plots represent the distribution of experimentally measured yield stresses for micropillars fabricated in matrix and twin grains. Individual micropillars examined by cross‑sectional lift‑out are indicated by single data points.} 
\label{fig.yieldstress}
\end{figure}

The yield stresses measured from 27 micropillars fabricated in both matrix and twin regions exhibit substantial scatter (Figure \ref{fig.yieldstress}). Although the pillars share the same \hkl[11-20] compression axis and comparable geometries, those fabricated in the twin region show a lower average yield stress and broader distribution. The interrupted experiments and atomistic simulations provide a framework for interpreting this variability in terms of the initial availability of twin embryos, mobile interfaces, and local stress concentrators. Depending on these initial conditions, the onset of twinning-related plasticity can involve the formation of a new embryo, expansion of an existing embryo, or migration of an established TB. 

The MD simulation results delineate how these different initial configurations affect the onset of plastic flow. Within the configurations examined, pillars containing pre-existing twin slabs with interfacial disconnections begin to thicken at lower applied stresses than those requiring the formation or expansion of a localized embryo. The onset stresses for newly forming and pre-existing embryos overlap, indicating that their relative values also depend on the local geometry and defect configuration. Complementary simulations using an alternative interatomic potential reproduce the qualitative contrast between the embryo and slab configurations (cf. Supplementary Figure S1). These comparisons support a qualitative interpretation of the experimental responses, whereas a quantitative correspondence between experimental and simulated stresses remains limited by the differences in specimen size, strain rate, and temperature, discussed in the following subsection. The interrupted experiments provide direct evidence that contrasting initial twin configurations are associated with different onset stresses. For instance, yielding in pillar TG105 at 81 MPa is associated with growth of a pre-existing twin, with migration of its boundary documented by the interrupted-test observations (Figure \ref{fig.TG105106}). In contrast, pillar MG007 reaches 225 MPa before the stress drop associated with the formation and rapid expansion of a new twin (Figure \ref{fig.MG007008}). In TG105, the existing interface provides a route for accommodating deformation without first creating a new twin embryo. This comparison supports the importance of initial interface availability in determining the measured yield stress. 

In a confined pillar, the first plastic event can be strongly influenced by the particular defects and interfaces retained within the sample volume.  When only a limited number of suitable activation sites are available, differences in their structure, position, and proximity to free surfaces can produce substantial variations in the nominal stress at which twinning begins. Pre-existing defects can facilitate embryo formation by generating local stress concentrations or providing favorable atomic configurations \cite{jeong2018situ, jiang2022visualization}. Their subsequent interactions with migrating interfaces can also impede growth, as illustrated by the obstacles encountered during embryo expansion in Figure \ref{fig.propagation}. The local defect configuration therefore influences both the availability of a twinning pathway and the conditions required to activate and sustain it. Twinning in such confined systems is therefore best viewed as a probabilistic event, controlled by the local defect landscape. 

Several sources of variability are relevant to the present experiments. As shown in Figure \ref{fig.nucleation}, the simulations demonstrate that differences in the contact geometry lead to pillar-to-pillar variations in local stress concentration, and hence in the activation stress, at which a stable twin first appears. Differences in pillar contact conditions could therefore contribute to the experimental stress distribution. In addition, EBSD mapping of the pre-deformed specimen, from which the micropillars were fabricated, reveals secondary twin variants with relatively small volumes introduced during channel-die compression (cf. Supplementary Figure S2). Although FIB milling targeted regions without surface-visible twins, this selection does not exclude subsurface twin lamellae. If such a lamella corresponds to the subsequently activated variant and its boundary is mobile, deformation can proceed through growth of the existing twin. Pillars containing such interfaces may consequently yield at lower stresses than pillars requiring the formation of a new embryo.

The lower average yield stress and broader distribution in the twin-grain pillars are consistent with differences in their prior deformation history. The KAM maps reveal local orientation variations within the previously twinned material (cf. Supplementary Figures S2 and S3), supporting the presence of microstructural heterogeneity. Favorable defect configurations or pre-existing mobile interfaces may facilitate twinning at lower applied stresses in some pillars, while differences in their availability, structure, and mobility can produce a broad range of activation conditions across the population. This variation provides a plausible microstructural basis for the more pronounced stochastic response of the twin-grain pillars. 

Together, these observations support a microstructural basis for the variability of twinning onset within confined volumes. A probabilistic description can represent differences between pillars in the availability of suitable activation sites and their effective activation thresholds. Heterogeneous twin activation also occurs in bulk Mg, where statistical studies show that grain orientation and size alone do not uniquely determine whether twinning occurs \cite{beyerlein2010statistical}. A more reproducible macroscopic response can emerge as deformation accumulates across many sites and through interactions between grains, even though individual twinning events remain sensitive to local conditions. The present results consequently motivate crystal plasticity descriptions that distinguish embryo formation from the growth of existing twins and account for the evolving availability of mobile interfaces. Incorporating distributions of effective activation thresholds would provide a means of connecting deformation history and local microstructure to the observed variability in twinning onset. 

\subsection{Cross-scale comparison between experiments and atomistic simulations}

The microcompression experiments and atomistic simulations employed in this study operate over substantially different length scales, temperatures, and strain rates. The experimental pillars have a diameter of approximately 4~$\mu$m and were compressed at room temperature at an approximate strain rate of $10^{-3}$~s$^{-1}$, whereas the simulated pillars have a diameter of 30~nm and were deformed at 10~K at a strain rate of $10^{8}$~s$^{-1}$. Consequently, a direct quantitative comparison of the activation stresses and the kinetics of twin nucleation, propagation, and thickening between the two approaches is not expected. Instead, the MD simulations are used as atomistic model systems to isolate individual stages of the twinning process that occur too rapidly and are too spatially localized to be resolved independently in the micropillar experiments. The meaningful comparison between the two approaches is therefore made at the level of deformation pathways, their sensitivity to the initial microstructural state, and the associated atomistic mechanisms.

The large difference in specimen size primarily affects the population of available defect sources and the role of free surfaces. A micrometer-sized experimental pillar may contain a heterogeneous population of potential sites for twin activation, including pre-existing dislocations and secondary twins from prior deformation, as well as defects introduced during FIB milling. In contrast, the initial defect population in the MD systems is well controlled and limited to the defects deliberately introduced into the starting configurations. Furthermore, owing to the large surface-to-volume ratio of the nanopillars, free-surface effects become increasingly important. Intrinsic surface stresses modify the local stress state, while traction-free surfaces and geometrical features such as edges and corners alter the energetics of defect nucleation and generate highly localized stress concentrations \cite{jennings2013modeling}. Such geometric effects can preferentially select heterogeneous nucleation sites and modify the corresponding activation stress, making twin nucleation particularly sensitive to the detailed surface morphology. This sensitivity is directly reflected in the present simulations (see Figure~\ref{fig.nucleation}), where changing the geometry of the truncated corner leads to a substantial change in the stress required to form a stable twin embryo. These configurations should therefore not be regarded as geometrically scaled replicas of the experimental pillars. Instead, they provide controlled realizations of different local nucleation environments and demonstrate the strong sensitivity of twin activation to the surface and local defect configuration. 

The strain rate difference between experiment and simulation introduces an additional limitation for comparison of absolute stresses. The MD strain rate of $10^{8}$~s$^{-1}$ is dictated by the accessible simulation time scale and exceeds the experimental strain rate by many orders of magnitude. Although deformation twinning does not require long-range atomic diffusion, the elementary events in twinning and competing dislocation activity are governed by thermally activated transitions across stress-dependent energy barriers. Increasing the applied stress lowers these activation barriers and thereby increases the probability of a transition within a given observation time. At the very high strain rates accessible to MD, the time available for thermal activation is substantially reduced, such that the applied stress must rise to higher values before the activation barrier becomes sufficiently small for the corresponding event to occur. This shifts the simulated response toward the stress-driven, near-athermal limit. The resulting activation stresses should therefore be interpreted as characteristic stresses under the MD conditions rather than as direct correlations with the experimental yield stress. Therefore, the objective of the simulations is to identify how the atomic-scale response changes when nucleation is required, when a twin embryo is already present, or when a mobile TB containing interfacial disconnections is available. Similar consideration applies to the difference in temperature. At 300~K, thermal fluctuations may modify the stability of small twin embryos and facilitate the migration of disconnections, thereby changing both the activation events and associated stresses \cite{xu2013importance,liu2019three}. Therefore, the simulations were performed at 10~K to suppress thermal fluctuations and provide a particularly clear view of the elementary atomic rearrangements associated with the advancing twin front and migrating TBs. 

Despite these differences, a meaningful correspondence emerges when experiments and simulations are compared in terms of the initial microstructural state and the deformation pathway that becomes available. Experimentally, e.g. pillar MG007, in which a new twin develops during loading, exhibits a pronounced high-stress event followed by rapid twin extension, whereas pillar TG105 already contains a secondary twin and deforms through migration of the pre-existing TB at substantially lower stress. The atomistic simulations provide a mechanistic analogue of this distinction: configurations requiring twin nucleation are strongly sensitive to the local surface and defect geometry, whereas systems containing a pre-existing embryo or TB can evolve without repeating the initial nucleation process, including through migration of interfacial disconnections. The simulations are therefore not used to quantitatively map MD activation stresses onto experimental yield stresses, but rather to identify the mechanisms accessible from different initial microstructural states. Beyond this mechanistic distinction, factors such as surface geometry, FIB-induced defects, and contact conditions may additionally contribute to the experimentally measured scatter.

\section{Conclusions}

In this study, \hkl{10-12} tensile twinning in confined volumes of Mg was systematically investigated by combining site-specific micropillar compression and interrupted microstructural characterization with MD simulations. The main conclusions are: 

\begin{itemize}
    \item Mg micropillars compressed along \hkl[11-20] direction deform primarily by \hkl{10-12} twinning. Discrete stress drops accompany twin formation and rapid propagation. Following lattice reorientation, subsequent deformation is mainly accommodated by basal slip within the twinned volume.
    \item MD simulations reveal rapid expansion of localized embryos through the advance of faceted interfaces, accompanied by stress relaxation. The pre-event stress peaks therefore characterize the onset of embryo expansion. Once forward advance is constrained by the pillar geometry, lateral thickening becomes predominant. In the pre-existing twin-slab configuration, migration of established TB proceeds through interfacial disconnection motion coupled with local atomic shuffling.
    \item The yield stresses of the 27 micropillars show substantial scatter, with twin-grain pillars exhibiting a lower mean and broader distribution than geometrically comparable matrix-grain pillars compressed along the same axis. This behavior is consistent with microstructural heterogeneity from prior deformation, where variations in embryo availability, interface mobility, and local stress concentrations govern the stochastic onset of twinning-related plasticity.
    \item   Although direct quantitative comparison between MD and micropillar stresses is limited by the differences in length scale, strain rate, and temperature, the simulations provide mechanistic insight into the deformation pathways accessible from different initial microstructural states. They show that local contact geometry, pre-existing embryos, and mobile TBs govern whether twinning initiates by nucleation, embryo propagation, or boundary migration, thereby linking deformation history to the stochastic first plastic event observed experimentally. 
  
\end{itemize}

\section*{Data Availability}
The data that support the findings of this study are available from the corresponding author upon reasonable request.

\section*{Acknowledgements}

H.W., T.A.S., and Z.X. acknowledge the financial support by the DFG (Grant Nr. 505716422). S.H.L., S.K.K., and Z.X. acknowledge financial support by the DFG through the SFB1394 Structural and Chemical Atomic Complexity – From Defect Phase Diagrams to Material Properties, project ID 409476157. Z.X. acknowledges financial support funded by the DFG – Projektnummer 562592407 and 555365333. F.Z.M. and J.G. acknowledge funding from the French National Research Agency (ANR), Grant ANR-21-CE08-0001 (ATOUUM) and ANR-22-CE92-0058-01 (SILA). The authors gratefully acknowledge the computing time provided to them at the NHR Center NHR4CES at RWTH Aachen University (project number p0020431). This is funded by the Federal Ministry of Education and Research, and the state governments participating on the basis of the resolutions of the GWK for national high performance computing at universities (www.nhr-verein.de/unsere-partner). The data used in this publication was managed using the research data management platform Coscine (http://doi.org/10.17616/R31NJNJZ) with storage space of the Research Data Storage (RDS) (DFG: INST222/1261-1) and DataStorage.nrw (DFG: INST222/1530-1) granted by the DFG and Ministry of
Culture and Science of the State of North Rhine-Westphalia.

\bibliographystyle{elsarticle-num}

\bibliography{main}

@article{wang2020effect,
  title={Effect of Al content on the critical resolved shear stress for twin nucleation and growth in Mg alloys},
  author={Wang, Jingya and Molina-Aldaregu{\'\i}a, Jon M and LLorca, Javier},
  journal={Acta Materialia},
  volume={188},
  pages={215--227},
  year={2020},
  publisher={Elsevier}
}

@article{pollock2010weight,
  title={Weight loss with magnesium alloys},
  author={Pollock, Tresa M},
  journal={Science},
  volume={328},
  number={5981},
  pages={986--987},
  year={2010},
  publisher={American Association for the Advancement of Science}
}

@article{molodov2022effect,
  title={Effect of gadolinium on the deformation and recrystallization behavior of magnesium crystals},
  author={Molodov, Konstantin D and Al-Samman, Talal and Molodov, Dmitri A},
  journal={Acta Materialia},
  volume={240},
  pages={118312},
  year={2022},
  publisher={Elsevier}
}

@article{lee2024atomistic,
  title={Atomistic investigation into the formation of axial weak twins during the compression of single-crystal Mg nanopillars},
  author={Lee, Jung Soo and Xu, Guanglong and Suh, Joung Sik and Bae, Jun Ho and Suh, Byeong-Chan and Kim, Young Min and Ko, Won-Seok and Lee, Byeong-Joo},
  journal={Acta Materialia},
  volume={263},
  pages={119512},
  year={2024},
  publisher={Elsevier}
}

@article{thompson2022lammps,
  title={LAMMPS-a flexible simulation tool for particle-based materials modeling at the atomic, meso, and continuum scales},
  author={Thompson, Aidan P and Aktulga, H Metin and Berger, Richard and Bolintineanu, Dan S and Brown, W Michael and Crozier, Paul S and In't Veld, Pieter J and Kohlmeyer, Axel and Moore, Stan G and Nguyen, Trung Dac and others},
  journal={Computer physics communications},
  volume={271},
  pages={108171},
  year={2022},
  publisher={Elsevier}
}

@article{kim2015modified,
  title={Modified embedded-atom method interatomic potentials for {Mg}--{X} ({X}= {Y}, {Sn}, {Ca}) binary systems},
  author={Kim, Ki-Hyun and Jeon, Jong Bae and Lee, Byeong-Joo},
  journal={Calphad},
  volume={48},
  pages={27--34},
  year={2015},
  publisher={Elsevier}
}

@article{wang2024defects,
  title={Defects in magnesium and its alloys by atomistic simulation: Assessment of semi-empirical potentials},
  author={Wang, Hexin and Gu{\'e}nol{\'e}, Julien and Korte-Kerzel, Sandra and Al-Samman, Talal and Xie, Zhuocheng},
  journal={Computational Materials Science},
  volume={240},
  pages={113025},
  year={2024},
  publisher={Elsevier}
}

@article{hirel2015atomsk,
  title={Atomsk: A tool for manipulating and converting atomic data files},
  author={Hirel, Pierre},
  journal={Computer Physics Communications},
  volume={197},
  pages={212--219},
  year={2015},
  publisher={Elsevier}
}

@article{stukowski2009visualization,
  title={Visualization and analysis of atomistic simulation data with {OVITO}--the {O}pen {V}isualization {T}ool},
  author={Stukowski, Alexander},
  journal={Modelling and simulation in materials science and engineering},
  volume={18},
  number={1},
  pages={015012},
  year={2009},
  publisher={IOP Publishing}
}

@article{jiang2022visualization,
  title={Visualization and validation of twin nucleation and early-stage growth in magnesium},
  author={Jiang, Lin and Gong, Mingyu and Wang, Jian and Pan, Zhiliang and Wang, Xin and Zhang, Dalong and Wang, Y Morris and Ciston, Jim and Minor, Andrew M and Xu, Mingjie and others},
  journal={Nature communications},
  volume={13},
  number={1},
  pages={20},
  year={2022},
  publisher={Nature Publishing Group UK London}
}

@article{xu2013importance,
  title = {On the importance of prismatic/basal interfaces in the growth of $(\overline{1}012)$ twins in hexagonal close‑packed crystals},
  author={Xu, Ben and Capolungo, Laurent and Rodney, David},
  journal={Scripta Materialia},
  volume={68},
  number={11},
  pages={901--904},
  year={2013},
  publisher={Elsevier}
}

@article{kiener2009micro,
  title={Micro-compression testing: A critical discussion of experimental constraints},
  author={Kiener, Daniel and Motz, Christian and Dehm, Gerhard},
  journal={Materials Science and Engineering: A},
  volume={505},
  number={1-2},
  pages={79--87},
  year={2009},
  publisher={Elsevier}
}

@article{conrad1957effect,
  title={Effect of temperature on the flow stress and strain-hardening coefficient of magnesium single crystals},
  author={Conrad, Hans and Robertson, WD},
  journal={Jom},
  volume={9},
  number={4},
  pages={503--512},
  year={1957},
  publisher={Springer}
}

@article{reed1957deformation,
  title={Deformation of magnesium single crystals by nonbasal slip},
  author={Reed-Hill, Robert E and Robertson, William D},
  journal={Jom},
  volume={9},
  number={4},
  pages={496--502},
  year={1957},
  publisher={Springer}
}

@article{flynn1961thermally,
  title={On the thermally activated mechanism of prismatic slip in magnesium single crystals},
  author={Flynn, Phillip Ward and Mote, JEDJ and Dorn, John Emil},
  journal={Transactions of the Metallurgical Society of AIME},
  volume={221},
  number={6},
  pages={1148--1154},
  year={1961}
}

@article{ando199211,
  title = {\{11--22\} $\langle \bar{1}\bar{1}23 \rangle$ Slip in Magnesium Single Crystal},
  author = {Ando, Shinji and Nakamura, Kanji and Takashima, Kazuki and Tonda, Hideki},
  journal = {Journal of Japan Institute of Light Metals},
  volume = {42},
  number = {12},
  pages = {765--771},
  year = {1992},
  doi = {10.2464/jilm.42.765}
}

@article{kelley1968plane,
  title={Plane-strain compression of magnesium and magnesium alloy crystals},
  author={Kelley, EW and Hosford, WFJR},
  journal={Trans Met Soc AIME},
  volume={242},
  number={1},
  pages={5--13},
  year={1968}
}

@article{christian1995deformation,
  title={Deformation twinning},
  author={Christian, John Wyrill and Mahajan, Subhash},
  journal={Progress in materials science},
  volume={39},
  number={1-2},
  pages={1--157},
  year={1995},
  publisher={Elsevier}
}

@article{nie2020microstructure,
  title={Microstructure, deformation, and property of wrought magnesium alloys},
  author={Nie, JF and Shin, KS and Zeng, ZR},
  journal={Metallurgical and Materials Transactions A},
  volume={51},
  number={12},
  pages={6045--6109},
  year={2020},
  publisher={Springer}
}

@article{tome2011multi,
  title={A multi-scale statistical study of twinning in magnesium},
  author={Tom{\'e}, CN and Beyerlein, IJ and Wang, J and McCabe, RJ},
  journal={Jom},
  volume={63},
  number={3},
  pages={19--23},
  year={2011},
  publisher={Springer}
}

@article{yoo1981slip,
  title={Slip, twinning, and fracture in hexagonal close-packed metals},
  author={Yoo, MH1641934},
  journal={Metallurgical transactions A},
  volume={12},
  number={3},
  pages={409--418},
  year={1981},
  publisher={Springer}
}

@article{godet2006use,
  title={Use of Schmid factors to select extension twin variants in extruded magnesium alloy tubes},
  author={Godet, St{\'e}phane and Jiang, Lin and Luo, AA and Jonas, JJ},
  journal={Scripta materialia},
  volume={55},
  number={11},
  pages={1055--1058},
  year={2006},
  publisher={Elsevier}
}

@article{chapuis2011temperature,
  title={Temperature dependency of slip and twinning in plane strain compressed magnesium single crystals},
  author={Chapuis, Adrien and Driver, Julian H},
  journal={Acta Materialia},
  volume={59},
  number={5},
  pages={1986--1994},
  year={2011},
  publisher={Elsevier}
}

@article{liu2017experimentally,
  title={Experimentally quantifying critical stresses associated with basal slip and twinning in magnesium using micropillars},
  author={Liu, Yue and Li, Nan and Kumar, M Arul and Pathak, S and Wang, J and McCabe, RJ and Mara, NA and Tom{\'e}, CN},
  journal={Acta Materialia},
  volume={135},
  pages={411--421},
  year={2017},
  publisher={Elsevier}
}

@article{beyerlein2010statistical,
  title={Statistical analyses of deformation twinning in magnesium},
  author={Beyerlein, IJ and Capolungo, L and Marshall, PE and McCabe, RJ and Tom{\'e}, CN},
  journal={Philosophical Magazine},
  volume={90},
  number={16},
  pages={2161--2190},
  year={2010},
  publisher={Taylor \& Francis}
}

@article{prasad2014micropillar,
  title={Micropillar and macropillar compression responses of magnesium single crystals oriented for single slip or extension twinning},
  author={Prasad, K Eswar and Rajesh, K and Ramamurty, U},
  journal={Acta materialia},
  volume={65},
  pages={316--325},
  year={2014},
  publisher={Elsevier}
}

@article{sim2019effect,
  title={Effect of temperature on the transition in deformation modes in Mg single crystals},
  author={Sim, Gi-Dong and Xie, Kelvin Y and Hemker, Kevin J and El-Awady, Jaafar A},
  journal={Acta Materialia},
  volume={178},
  pages={241--248},
  year={2019},
  publisher={Elsevier}
}

@article{sim2018anomalous,
  title={Anomalous hardening in magnesium driven by a size-dependent transition in deformation modes},
  author={Sim, Gi-Dong and Kim, Gyuseok and Lavenstein, Steven and Hamza, Mohamed H and Fan, Haidong and El-Awady, Jaafar A},
  journal={Acta Materialia},
  volume={144},
  pages={11--20},
  year={2018},
  publisher={Elsevier}
}

@article{jeong2018situ,
  title={In-situ TEM observation of $\{10\overline{1}2\}$ twin-dominated deformation of {Mg} pillars: Twinning mechanism, size effects and rate dependency},
  author={Jeong, Jiwon and Alfreider, Markus and Konetschnik, Ruth and Kiener, Daniel and Oh, Sang Ho},
  journal={Acta Materialia},
  volume={158},
  pages={407--421},
  year={2018},
  publisher={Elsevier}
}

@article{della2023micromechanical,
  title={Micromechanical response of pure magnesium at different strain rate and temperature conditions: twin to slip and slip to twin transitions},
  author={della Ventura, Nicolo Maria and Schweizer, Peter and Sharma, Amit and Jain, Manish and Edwards, Thomas Edward James and Schwiedrzik, J Jakob and Peruzzi, Cinzia and Loge, Roland E and Michler, Johann and Maeder, Xavier},
  journal={Acta Materialia},
  volume={243},
  pages={118528},
  year={2023},
  publisher={Elsevier}
}

@article{della2023temperature,
  title={Temperature dependent critical stress for $\{10\overline{1}2\}$ twinning in magnesium micropillars at cryogenic temperatures},
  author={della Ventura, Nicolo M and Tian, Chunhua and Sharma, Amit and Edwards, Thomas EJ and Schwiedrzik, J Jakob and Loge, Roland E and Michler, Johann and Maeder, Xavier},
  journal={Scripta Materialia},
  volume={226},
  pages={115195},
  year={2023},
  publisher={Elsevier}
}

@article{mathis2021dynamics,
  title={On the dynamics of twinning in magnesium micropillars},
  author={M{\'a}this, Kristi{\'a}n and Knapek, Michal and {\v{S}}i{\v{s}}ka, Filip and Harcuba, Petr and Ugi, D{\'a}vid and Isp{\'a}novity, P{\'e}ter Dus{\'a}n and Groma, Istv{\'a}n and Shin, Kwang Seon},
  journal={Materials \& Design},
  volume={203},
  pages={109563},
  year={2021},
  publisher={Elsevier}
}

@article{della2021101,
  title={$\{10\overline{1}2\}$ twinning mechanism during in situ micro-tensile loading of pure Mg: Role of basal slip and twin-twin interactions},
  author={Della Ventura, Nicolo M and Kalacska, Szilvia and Casari, Daniele and Edwards, Thomas EJ and Sharma, Amit and Michler, Johann and Loge, Roland and Maeder, Xavier},
  journal={Materials \& Design},
  volume={197},
  pages={109206},
  year={2021},
  publisher={Elsevier}
}

@phdthesis{kim2011small,
  title={Small volume investigation of slip and twinning in magnesium single crystals},
  author={Kim, Gyu Seok},
  year={2011},
  school={Grenoble}
}

@article{serra1996new,
  title={A new model for $\{10\overline{1}2\}$ twin growth in hcp metals},
  author={Serra, A and Bacon, DJ},
  journal={Philosophical Magazine A},
  volume={73},
  number={2},
  pages={333--343},
  year={1996},
  publisher={Taylor \& Francis}
}

@article{wang20091,
  title={$(\overline{1}012)$ Twinning nucleation mechanisms in hexagonal-close-packed crystals},
  author={Wang, Jian and Hirth, JP and Tom{\'e}, CN},
  journal={Acta Materialia},
  volume={57},
  number={18},
  pages={5521--5530},
  year={2009},
  publisher={Elsevier}
}

@article{liu2019three,
  title={Three-dimensional character of the deformation twin in magnesium},
  author={Liu, Y and Tang, PZ and Gong, MY and McCabe, RJ and Wang, J and Tom{\'e}, CN},
  journal={Nature communications},
  volume={10},
  number={1},
  pages={3308},
  year={2019},
  publisher={Nature Publishing Group UK London}
}

@article{xie2021twin,
  title={Twin boundary migration mechanisms in quasi-statically compressed and plate-impacted Mg single crystals},
  author={Xie, Kelvin Y and Hazeli, Kavan and Dixit, Neha and Ma, Luoning and Ramesh, KT and Hemker, Kevin J},
  journal={Science Advances},
  volume={7},
  number={42},
  pages={eabg3443},
  year={2021},
  publisher={American Association for the Advancement of Science}
}

@article{larsen2016robust,
  title={Robust structural identification via polyhedral template matching},
  author={Larsen, Peter Mahler and Schmidt, S{\o}ren and Schi{\o}tz, Jakob},
  journal={Modelling and Simulation in Materials Science and Engineering},
  volume={24},
  number={5},
  pages={055007},
  year={2016},
  publisher={IOP Publishing}
}

@article{honeycutt1987molecular,
  title={Molecular dynamics study of melting and freezing of small Lennard-Jones clusters},
  author={Honeycutt, J Dana and Andersen, Hans C},
  journal={Journal of Physical Chemistry},
  volume={91},
  number={19},
  pages={4950--4963},
  year={1987},
  publisher={ACS Publications}
}

@article{jeong2025nanoscale,
  title={Nanoscale mechanisms limiting non-basal plasticity in magnesium},
  author={Jeong, Jiwon and Xie, Zhuocheng and Alfreider, Markus and Korte-Kerzel, Sandra and Kiener, Daniel and Gu{\'e}nol{\'e}, Julien and Oh, Sang Ho},
  journal={Acta Materialia},
  volume={296},
  pages={121261},
  year={2025},
  publisher={Elsevier}
}

@article{xie2025atomic,
  title={Atomic-scale modeling of defects in magnesium and its alloys: A review},
  author={Xie, Zhuocheng and Gu{\'e}nol{\'e}, Julien and Wang, Hexin and Petrazoller, Jo{\'e} and Mouhib, Fatim-Zahra and Guitton, Antoine and Richeton, Thiebaud and Berbenni, St{\'e}phane and Al-Samman, Talal},
  journal={Journal of Magnesium and Alloys},
  year={2025},
  publisher={Elsevier}
}

@article{ovri2023mechanistic,
  title={Mechanistic origin of the enhanced strength and ductility in Mg-rare earth alloys},
  author={Ovri, Henry and Markmann, J{\"u}rgen and Barthel, Juri and Kruth, Maximilian and Dieringa, Hajo and Lilleodden, Erica T},
  journal={Acta Materialia},
  volume={244},
  pages={118550},
  year={2023},
  publisher={Elsevier}
}

@article{uchic2004sample,
  title={Sample dimensions influence strength and crystal plasticity},
  author={Uchic, Michael D and Dimiduk, Dennis M and Florando, Jeffrey N and Nix, William D},
  journal={Science},
  volume={305},
  number={5686},
  pages={986--989},
  year={2004},
  publisher={American Association for the Advancement of Science}
}

@article{yu2010strong,
  title={Strong crystal size effect on deformation twinning},
  author={Yu, Qian and Shan, Zhi-Wei and Li, Ju and Huang, Xiaoxu and Xiao, Lin and Sun, Jun and Ma, Evan},
  journal={nature},
  volume={463},
  number={7279},
  pages={335--338},
  year={2010},
  publisher={Nature Publishing Group UK London}
}

@article{bei2008effects,
  title={Effects of pre-strain on the compressive stress--strain response of {Mo}-alloy single-crystal micropillars},
  author={Bei, Hongbin and Shim, Sanghoon and Pharr, George Mathews and George, Easo P},
  journal={Acta Materialia},
  volume={56},
  number={17},
  pages={4762--4770},
  year={2008},
  publisher={Elsevier}
}

@article{kraft2010plasticity,
  title={Plasticity in confined dimensions},
  author={Kraft, Oliver and Gruber, Patric A and M{\"o}nig, Reiner and Weygand, Daniel},
  journal={Annual review of materials research},
  volume={40},
  number={1},
  pages={293--317},
  year={2010},
  publisher={Annual Reviews}
}

@article{greer2011plasticity,
  title={Plasticity in small-sized metallic systems: Intrinsic versus extrinsic size effect},
  author={Greer, Julia R and De Hosson, Jeff Th M},
  journal={Progress in Materials Science},
  volume={56},
  number={6},
  pages={654--724},
  year={2011},
  publisher={Elsevier}
}

@article{hoover1985canonical,
  title={Canonical dynamics: {E}quilibrium phase-space distributions},
  author={Hoover, William G},
  journal={Physical review A},
  volume={31},
  number={3},
  pages={1695},
  year={1985},
  publisher={APS}
}

@article{bitzek2006structural,
  title={Structural relaxation made simple},
  author={Bitzek, Erik and Koskinen, Pekka and G{\"a}hler, Franz and Moseler, Michael and Gumbsch, Peter},
  journal={Physical review letters},
  volume={97},
  number={17},
  pages={170201},
  year={2006},
  publisher={APS}
}

@article{guenole2020assessment,
  title={Assessment and optimization of the fast inertial relaxation engine (fire) for energy minimization in atomistic simulations and its implementation in lammps},
  author={Gu{\'e}nol{\'e}, Julien and N{\"o}hring, Wolfram G and Vaid, Aviral and Houll{\'e}, Fr{\'e}d{\'e}ric and Xie, Zhuocheng and Prakash, Aruna and Bitzek, Erik},
  journal={Computational Materials Science},
  volume={175},
  pages={109584},
  year={2020},
  publisher={Elsevier}
}

@article{jennings2013modeling,
  title={Modeling dislocation nucleation strengths in pristine metallic nanowires under experimental conditions},
  author={Jennings, Andrew T and Weinberger, Christopher R and Lee, Seok-Woo and Aitken, Zachary H and Meza, Lucas and Greer, Julia R},
  journal={Acta materialia},
  volume={61},
  number={6},
  pages={2244--2259},
  year={2013},
  publisher={Elsevier}
}

@article{pond2016atomic,
  title={Atomic displacements accompanying deformation twinning: shears and shuffles},
  author={Pond, Robert C and Hirth, JP and Serra, A and Bacon, DJ},
  journal={Materials Research Letters},
  volume={4},
  number={4},
  pages={185--190},
  year={2016},
  publisher={Taylor \& Francis}
}

\end{document}